\documentclass[12pt,english]{article}
\usepackage[T1]{fontenc}
\usepackage[utf8]{inputenc}
\usepackage{geometry}
\usepackage{color}
\usepackage{float}
\usepackage{multirow}
\usepackage{amsmath,bm}
\usepackage{amssymb}
\usepackage{graphicx}
\usepackage{setspace}
\usepackage[authoryear]{natbib}
\usepackage{tikz}
\usepackage{adjustbox}
\usetikzlibrary{arrows,shapes,automata,backgrounds,decorations,petri,positioning,patterns}
\makeatletter

\providecommand{\tabularnewline}{\\}

\usepackage[T1]{fontenc}
\usepackage[utf8]{inputenc}
\usepackage{geometry}
\usepackage{color}
\usepackage{float}
\usepackage{multirow}
\usepackage{amsmath}
\usepackage{amssymb}
\usepackage{graphicx}
\usepackage{setspace}
\usepackage[authoryear]{natbib}
\usepackage{tikz}
\usepackage{adjustbox}

\usepackage{amsthm}\usepackage{physics}
\usepackage{gensymb}
\usepackage{flexisym}
\usepackage{tikz}
\usepackage{mathrsfs}\usepackage{rotating}\usepackage{titlesec}
\usepackage{lipsum}\usepackage{fancyhdr}
\usepackage{framed}\usepackage{chngcntr}\usepackage{blindtext}
\usepackage{epstopdf}
\usepackage{multirow}
\usepackage{colortbl}
\usepackage{pdflscape}
\usepackage{bbm}
\usepackage{xcolor}

\newcommand{\btheta}{\boldsymbol \theta}

\newcommand{\bfbeta}{\boldsymbol \beta}

\newcommand{\bfeta}{\boldsymbol \eta}

\newcommand{\bftheta}{\boldsymbol \theta}
\newcommand{\bfmu}{\boldsymbol \mu}
\newcommand{\bfphi}{\boldsymbol \phi}

\newcommand{\bfgamma}{\boldsymbol \gamma}
\newcommand{\bfsigma}{\boldsymbol \sigma}
\newcommand{\bfSigma}{\boldsymbol \Sigma}

\newcommand{\bfLambda}{\boldsymbol \Lambda}

\newcommand{\bfTheta}{\boldsymbol \Theta}
\newcommand{\bfS}{\mathbf S}
\newcommand{\bfy}{\mathbf y}
\newcommand{\bfZ}{\mathbf Z}
\newcommand{\bfw}{\mathbf w}

\newcommand{\bs}{\mathbf s}

\newcommand{\bfC}{\mathbf C}
\newcommand{\bfA}{\mathbf A}
\newcommand{\bfD}{\mathbf D}

\numberwithin{equation}{section}

\providecommand{\tabularnewline}{\\}
\floatstyle{ruled}
\newfloat{algorithm}{tbp}{loa}
\providecommand{\algorithmname}{Algorithm}
\floatname{algorithm}{\protect\algorithmname}

\newcommand{\distas}[1]{\mathbin{\overset{#1}{\kern\z@\sim}}}%
\newsavebox{\mybox}\newsavebox{\mysim}
\newcommand{\distras}[1]{%
  \savebox{\mybox}{\hbox{\kern3pt\scriptstyle#1\kern3pt}}%
  \savebox{\mysim}{\hbox{\sim}}%
  \mathbin{\overset{#1}{\kern\z@\resizebox{\wd\mybox}{\ht\mysim}{\sim}}}%
}

\usepackage{geometry}

\usepackage{setspace}
\@ifundefined{showcaptionsetup}{}{%
 \PassOptionsToPackage{caption=false}{subfig}}
\usepackage{subfig}

\makeatother

\usepackage{authblk}
\defcitealias{NOAA}{NOAA2019}

\newcommand{\blind}{1}

\usepackage{babel}
\begin{document}

\begin{singlespace}
\if1\blind {
\title{Recursive Nearest Neighbor Co-Kriging Models for Big Multi-fidelity Spatial Data Sets
}
\author[1]{Si Cheng}
 \author[1]{Bledar A. Konomi  \thanks{Corresponding author:Bledar A. Konomi (alex.konomi@uc.edu)} }
\author[2]{Georgios Karagiannis}
\author[1]{Emily L. Kang}
\affil[1]{Division of Statistics and Data Sciences, Department of Mathematical Sciences, University of Cincinnati, USA}
\affil[2]{Department of Mathematical Sciences, Durham University, UK}
\maketitle 
} \fi
\end{singlespace}

\begin{abstract}
\begin{singlespace}

Big datasets are gathered daily from different remote sensing platforms.  Recently, statistical co-kriging models, with the help of  scalable techniques, have been able to combine such datasets by using  spatially varying bias corrections.  The associated Bayesian inference for these models is usually facilitated via Markov chain Monte Carlo (MCMC) methods which present (sometimes prohibitively) slow mixing and convergence because they require the simulation of high-dimensional random effect vectors from their posteriors given large datasets.  To enable fast inference in big data spatial problems, we propose the recursive nearest neighbor co-kriging (RNNC) model. Based on this model, we develop two computationally efficient inferential procedures: a) the collapsed  RNNC which reduces the posterior sampling space by integrating out the latent processes, and b) the  conjugate RNNC, an MCMC free inference which significantly reduces the computational time without sacrificing prediction accuracy. An important highlight of conjugate RNNC is that it enables fast inference in massive multifidelity data sets by avoiding expensive integration algorithms. The efficient computational and good predictive performances of our proposed algorithms are demonstrated on benchmark examples and the analysis of the High-resolution Infrared Radiation Sounder data gathered from two NOAA polar orbiting satellites in which we managed to reduce the computational time from multiple hours to just a few minutes.

\end{singlespace}
\end{abstract}
\begin{singlespace}
\end{singlespace}


\section{Introduction}

Global geophysical information is measured daily by  numerous satellite sensors. Due to aging and exposure to the harsh environment of space the satellite sensors degrade over time, resulting in decreased performance reliability. Decreased performance may affect data measurement accuracy\citep{goldberg2011}. In addition, newer satellites  with technologically more advanced sensors provide information of higher fidelity than older sensors. These  discrepancies in sensor performance have created the need to develop efficient methods to analyse daily  global remote sensing measurements with varying fidelity. Here, our work is motivated by a data set produced from the high-resolution infrared radiation sounder (HIRS), which provides  hundred of thousands of measurements from multiple satellite platforms daily.

Multiple methods in remote sensing have been developed to assess satellite sensor performance and consistency \citep{chander2013, xiong2010, nrc2004}. These  methods do not account for spatial correlation and oversimplify the relationship between sensors. Furthermore, statistical methods to analyse these data sets which account for spatial correlation pose challenges due to the multifideltiy presence as well as the size and computationally intensive procedures. \citet{nguyen2012spatial,nguyen2017multivariate} have proposed data fusion techniques to model multivariate spatial data at potentially different spatial resolutions based on fixed ranked kriging \citep{cressie2008fixed}. The accuracy of this approach relies on the number of basis functions
and can only capture large scale variation of the covariance function. When the data sets are dense, strongly correlated, and the noise effect is sufficiently small, the low rank kriging techniques have difficulty accounting for small scale variation \citep{stein2014limitations}.

Many statistical methods for large spatially correlated data sets have been developed over the past two decade.  For instance, we distinguish, the low-rank approximation methods \citep{banerjee2008gaussian,cressie2008fixed}, approximate likelihood methods \citep{stein2004approximating,Gramacy2015}, covariance tapering methods \citep{furrer2006covariance,kaufman2008covariance,du2009fixed}, sparse structures \citep{lindgren2011explicit,nychka2015multiresolution,datta2016hierarchical, MaKang2020,MeshedGP2020}, lower dimensional conditional distributions \citep{vecchia1988estimation,stein2004approximating,datta2016hierarchical,katzfuss2017general}, and multiple-scale approximation \citep{Sang2012, Katzfuss2016, Abdulah2023,Shirota_Finley2022}. All these methods have been developed for  data sets obtained from the same source or instrument which translates to a single fidelity data source. However, their extension to multi-fidelity data sets is not  straightforward.

Autoregressive co-kriging models \citep{kennedy2000predicting,ZhiguangConnerJanetWu2005,le2013bayesian}, originally built for computer simulation problems, can be used for the analysis of multiple fidelity remote sensing observations with spatially nested structure and no random error. Nested design in the multifidelity setting means the design points at the higher fidelity levels are subsets of the lower fidelity ones. \cite{konomikaragiannisABTCK2019} and \cite{ma2019multifidelity} relaxed the nested design requirements by properly introducing an imputation mechanism. However, the aforesaid methods rely on Gaussian process models and are computationally  impossible  for big data problems. For cases when the observed space can be expressed as a tensor  product \citet{Konomi2022_JABES} uses a separable covariance function within the co-kriging model to improve the computational efficiency.  For large data sets which are irregularly positioned over space and contaminated with random error, \citet{Si_etall2020_NNCGP} proposed the nearest neighbour co-kriging Gaussian process (NNCGP) to embed  nearest neighbor Gaussian process \citep[NNGP;][]{datta2016hierarchical} into an autoregressive co-kriging model to make computations possible. NNCGP achieves this by using imputation ideas into the latent variables to construct a nested reference set of multiple NNGP levels. Although NNCGP makes the analysis of big multi-fidelity data sets computationally possible,  its computational speed depends on an expensive iterative MCMC procedure which makes it impractical for analysing daily large data sets. 

To overcome the iterative MCMC procedure, we propose a recursive formulation based on the latent variable of the NNCGP model following similar ideas with  \citet{le2014recursive}, who proposed the recursive formulation directly into the observations.  Based on this new formulation, which we call recursive nearest neighbors co-kriging (RNNC), we are able to build a nearest neighbors co-kriging model with $T$ levels by building $T$ conditionally independent NNGPs. This enables the development of two alternative inferential procedures which aim to reduce high-dimensional parametric space, improve convergence, and reduce computational time in comparison to the NNCGP. Both proposed procedures are able to address applications for large non-nested and irregular spatial data sets from different platforms and with varying quality. The first proposed procedure, called  Collapsed RNNC,  reduces the MCMC posterior sampling space by integrating out the spatial latent variables. Based on the collapsed RNNC, we propose an MCMC free procedure to speed up Bayesian inference. We build an algorithm which sequentially decomposes the parametric space into conditionally independent parts for each fidelity level where we can apply a K-fold optimization method. Each sequential step can be viewed as collapsed NNGP \citep{finley2019JCGS} where the bases functions of the Gaussian process mean are determined at the previous step. We name this second inferential procedure conjugate RNNC.  We note that the MCMC free procedure proposed in \citep{finley2019JCGS} cannot be applied directly in the NNCGP model because the computational complexity of the $K$-fold cross-validation method depends on the dimension of the parametric space. Our simulation study and our analysis of the HERS data set  shows that the proposed conjugate RNNC procedure reduces the computational time notably without significantly sacrificing prediction accuracy over the existing NNCGP approach. 

The layout of the paper is as follows. In  Section~\ref{sec: data}, we introduce the high-resolution infrared radiation sounder data  studied in this work. In Section~3, we review the NNCGP model. In Section~4, we introduce the proposed RNNC model. In Section~4.1, we integrate out the latent variables from the model and design an MCMC algorithm for this model. In Section~4.2, we design an MCMC free approach tailored to the proposed RNNC model that facilitates parametric and predictive inference. In Section~5, we investigate the performance of the proposed procedure. Specifically in Section~5.1 we introduce two simulation studies and in Section~5.2 we implement the proposed method for the analysis of data sets from two satellites, NOAA-14 and NOAA-15. Finally,in Section~6 we give a  summary and conclusion.

\section{High-resolution Infrared Radiation Sounder Data}\label{sec: data}

Satellite soundings have been providing measurements of the Earth’s atmosphere, oceans, land, and ice since the 1970s to support the study of global climate system dynamics. Long term observations from past
and current environmental satellites are widely used in developing
climate data records (CDR) \citep{nrc2004}. 
HIRS mission objectives include observations of atmospheric temperature,
water vapor, specific humidity, sea surface temperature, cloud cover,
and total column ozone. The HIRS instrument is comprised of twenty
channels, including twelve longwave channels, seven shortwave channels,
and one visible channel. 
The dataset being considered in this study is limb-corrected
HIRS swath data as brightness temperatures \citep{jackson2003}. The
data is stored as daily files, where each daily file records approximately
120,000 geolocated observations. The current archive includes data from NOAA-5 through NOAA-17 along with Metop-02, covering the time period of 1978-2017. In all, this data archive is more than 2 TB, with an average daily file size of about 82 MB.  The  HIRS CRD faces some common challenges regarding the consistency and accuracy over time, due to degradation of sensors and intersatellite discrepancies. Furthermore, there is missing information caused by atmospheric conditions such as thick cloud cover.

\begin{figure}
\centering
\subfloat[{{{\large{}\label{fig:sfig1-1} }Observations
of NOAA 14}}]{\includegraphics[width=0.49\linewidth]
{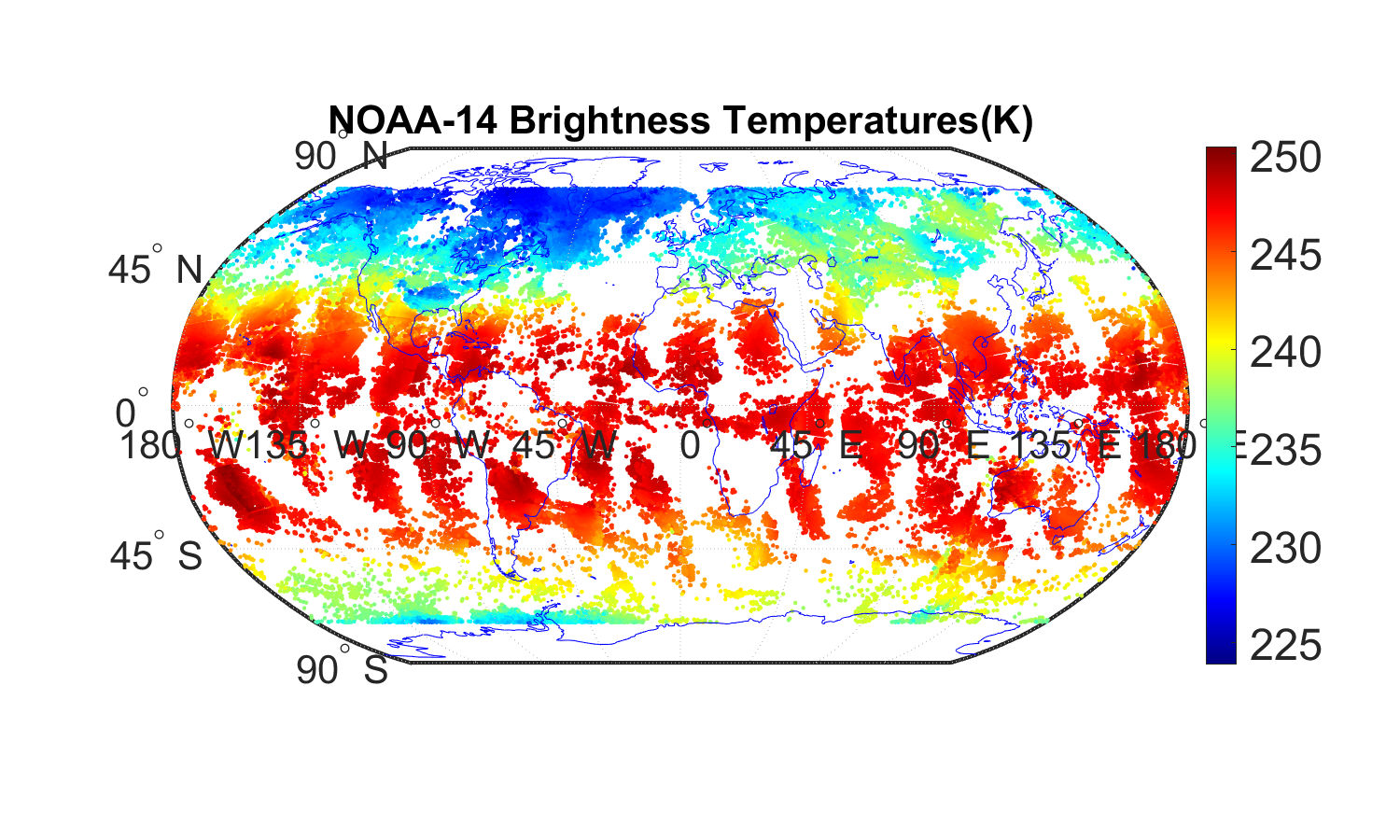}}\hfill
\subfloat[{{{\large{}\label{fig:sfig2-1}}Training
data of NOAA 15}}]{{}{}\includegraphics[width=0.49\linewidth]{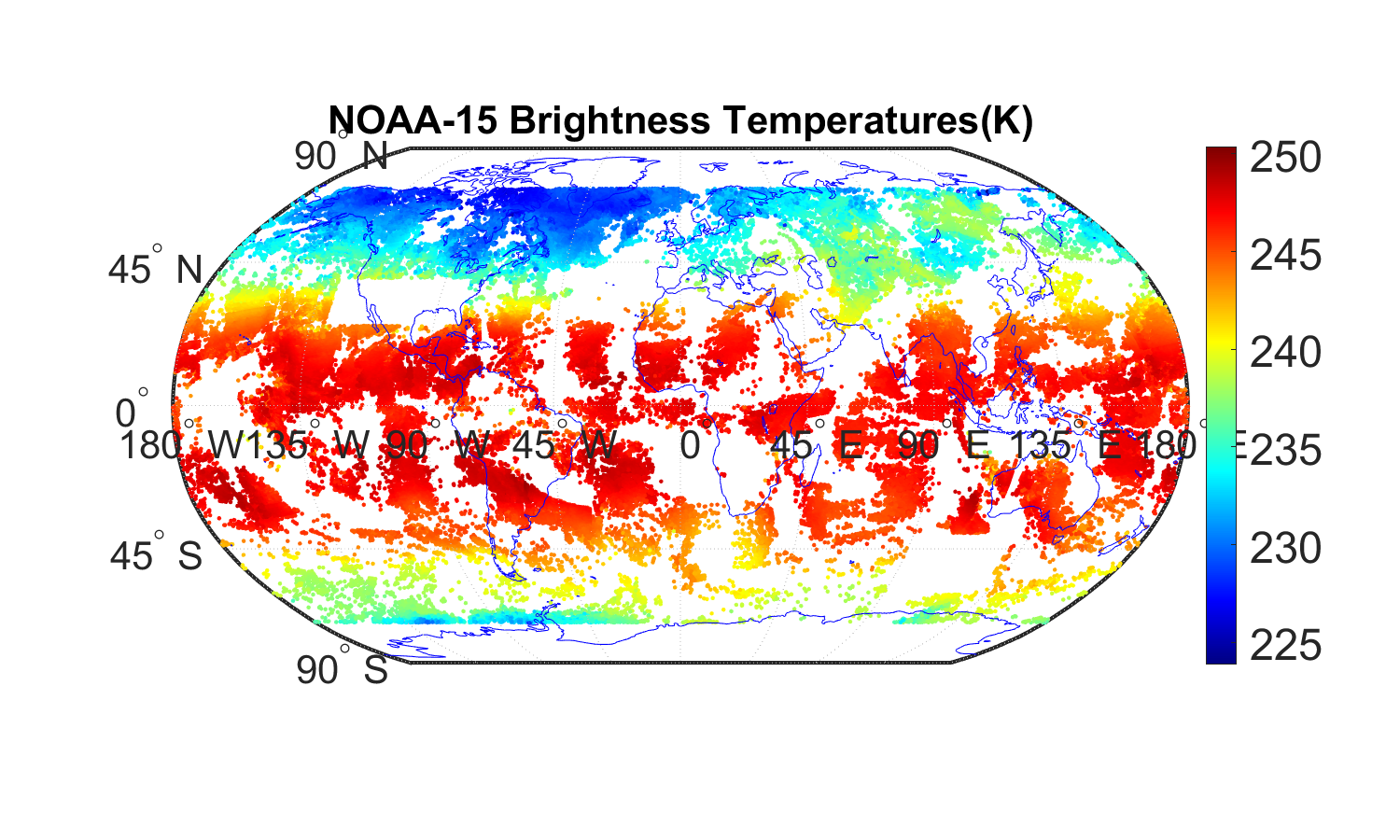}}{}{} 
\caption{NOAA-14 Brightness
Temperatures observation data-set, NOAA-15 Brightness Temperatures
training data-set for Channel 5 on March 1, 2001.{\large\label{fig:fig-1}}}
\end{figure}

We examine HIRS Channel 5 observations from a single day, March
1, 2001, as illustrated in Figure \ref{fig:fig-1}. On this day, we may exploit a period of temporal overlap in the NOAA POES series where two satellites captured measurements: NOAA-14 and NOAA-15. The HIRS sensors on these two satellites have similar technical designs which allow us to ignore the spectral and spatial footprint differences. NOAA-14 became operational in December 1994 while NOAA-15 became operational in October 1998. The spatial resolution footprint for both satellites is  approximately 10 km at nadir. Given the sensor age difference, it is reasonable to consider that the instruments on-board NOAA-15 are in better condition than those of NOAA-14 and hence provide more accurate data. Therefore, we treat observations from NOAA-14 as a low fidelity dataset, and those from NOAA-15 as a high fidelity dataset.

\section{Nearest Neighbor Co-kriging Gaussian Process}\label{sec:sequentialNNCGP}
Let $y_{t}(\bs)$ denote the output function at the spatial location
$s$ at fidelity level $t=1,...,T$ in a system with fidelity $T$ levels. The fidelity level index $t$ runs from the least accurate to the most accurate one. Let $z_{t}(\bs)$  denote the observed output at location $s$. We specify the co-kriging model as: 
\setlength{\belowdisplayskip}{5pt}\setlength{\belowdisplayshortskip}{5pt}
\setlength{\abovedisplayskip}{5pt}\setlength{\abovedisplayshortskip}{5pt}
\begin{align}
 & z_{t}(\bs)=y_{t}(\bs)+\epsilon_{t},\nonumber \\
 & y_{t}(\bs)=\zeta_{t-1}(\bs)y_{t-1}(\bs)+\delta_{t}(\bs),\label{eq:davdsgdaf}\\
 & \delta_{t}(\bs)=\mathbf{h}^T_{t}(\bs)\bfbeta_{t}+w_{t}(\bs), \nonumber 
\end{align}
where $z_{t}(\bs)$ is contaminated by additive random noise $\epsilon_{t}\sim N(0,\tau_{t}^2)$ for $t=2,\ldots,T$, and $y_{1}(\bs)=\mathbf{h}^T_{1}(\bs)\boldsymbol{\beta_{1}}+w_{1}(\bs)$ is the noiseless output. 
Here, $\zeta_{t-1}(\bs)$
and $\delta_{t}(\bs)$ represent the scale and additive discrepancies
between systems with fidelity levels $t$ and $t-1$, $\mathbf{h}_{t}(\cdot)$ is a vector of preselected bases functions, and $\bfbeta_{t}$
is a vector of coefficients at fidelity level $t$. 
The latent random function $w_{t}(\bs)$ is modeled as a Gaussian process, mutually independent for different $t$; i.e. $w_{t}(\cdot)\sim GP(0,C_{t}(\cdot,\cdot;\btheta_{t}))$
where $C_{t}(\cdot,\cdot;\btheta_{t})$ is a covariance
function with covariance parameters $\btheta_{t}$ at fidelity 
level $t$.  Any well defined covariance function can be used  $C_t(\bs,\bs'|\boldsymbol{\theta}_t)=\sigma_t^{2}R(\bs,\bs'|\boldsymbol{\phi}_t)$,
where $\boldsymbol{\theta}_t=\{\sigma_t^2,\boldsymbol{\phi}_t\}$. This indicates that discrepancy term $\delta_{t}(\bs)$ is a Gaussian process.  Finally, the unknown scale discrepancy function $\zeta_{t-1}(\bs)$ is modeled as a basis expansion  $\zeta_{t-1}(\bs|\bfgamma_{t-1})=\mathbf{g}_{t-1}(\bs)^{T}\bfgamma_{t-1}$ (usually low degree), 
where $\mathbf{g}_{t}(\bs)$ is a vector of polynomial basis functions and $\{\bfgamma_{t-1}\}$ is a vector of 
random coefficients, for $t=2,\dots,T$.

Let us assume the system is observed at $n_t$ locations at  fidelity level $t$. Let $\bfS_{t}=\{\bs_{t,1},\dots,\bs_{t,n_t}\}$ be the set of $n_t$ observed locations, let $\bfw_t={w}_{t}(\bfS_t)=\{w_{t}(\bs_{t,1}),\dots,w_{t}(\bs_{t,n_{t}})\}$ the latent spatial random effect vector at fidelity level $t$, and let $\bfZ_{t}={z}_{t}(\bfS_t)=\{z_{t}(\bs_{t,1}),\dots,z_{t}(\bs_{t,n_{t}})\}$ represent
the observed output at fidelity level $t$. If data $\{\bfZ_{t}\}$ are observed in non-nested  locations across the fidelity levels, the calculation of the likelihood requires $\mathcal{O}((\sum_{t=1}^{T}n_t)^3)$ flops to invert the covariance matrix of the observations (denoted by $\boldsymbol{\Lambda}$) and additional $\mathcal{O}((\sum_{t=1}^{T}n_t)^2)$ memory to store it as explained in \citep{konomikaragiannisABTCK2019}. To reduce the computational complexity, \citet{Si_etall2020_NNCGP} proposed NNCGP which assigns conditionally independent   NNGP models within a nested reference set.  For Bayesian inference, \citet{Si_etall2020_NNCGP} proposed a Gibbs sampler taking advantage of the nearest neighbour structure at each fidelity level. However, this sampler is based on updating a conditionally independent high dimensional latent variable which  could cause  slow convergence and high autocorrelation \citep{LiuEtAl1994}. The slow convergence can  significantly increase the number of the Gibbs sampler iterations $I$. The overall computational cost of NNCGP for $m$ neighbours and non-nested spatial locations is $\mathcal{O}(I\times(\sum_{t=1}^{T}n_{t})m^{3})$ floating point operations (flops). 
Also, for a fixed computational budget, the produced Monte Carlo estimates may be sensitive to the initial values of the MCMC sampler. Despite reducing the computational  complexity to linear for every MCMC iteration, the number of the iterations can significantly increase computational cost.  This simple observation makes the existing NNCGP too expensive for the vast majority of real remote sensing applications.

\section{Recursive Nearest Neighbor Co-kriging Model}

Improvement in the convergence of MCMC  can be achieved by integrating out the latent variable $\bfw=(\bfw_1, \dots, \bfw_{S})$ from the Bayesian hierarchical NNCGP model which allows dimension reduction in the sampling space and the involved posterior distributions. However, integrating out the latent variables $\bfw$ in the NNCGP model is not feasible under non-nested designs. This is because the posterior distribution of latent variable of the lower fidelity is affected by the likelihood of higher fidelity. To make possible the integration of latent variables $\bfw$,  we propose a recursive formulation for the co-kriging model by using ideas similar to \citep{le2014recursive}.  Precisely, our proposed recursive nearest neighbors co-kriging (RNNC) has the following hierarchical structure: 
\begin{align}
 z_{t}(\bs)&=y_{t}(\bs)+\epsilon_{t}\nonumber \\
  y_{t}(\bs)& =\zeta_{t-1}(\bs)\hat{y}_{t-1}(\bs)+\delta_{t}(\bs),\label{eq:recursive}\\
 \delta_{t}(\bs) & =\mathbf{h}^T_{t}(\bs)\bfbeta_{t}+w_{t}(\bs), \nonumber 
\end{align}
where $\delta_{t}(\bs)$ is a Gaussian process as before and   $\hat{y}_{t-1}(\bs)$ is a Gaussian process with distribution $[\hat{y}_{t-1}(\bs)|\bfZ_{t-1},\hat{y}_{t-2}(\bs), \bftheta_{t-1},\bfbeta_{t-1}]$. Essentially, we express $y_t(\bs)$ (the Gaussian process response at level $t$) as a function of the Gaussian process
$y_{t-1}(\bs)$ conditioned by the values $\bfZ^{(t-1)} = (\bfZ_1,\dots, \bfZ_{t-1})$.  For computational efficiency,  we assume NNGP independent priors for  $w_{t}(\bs)$, $t=1,\dots, T$. Based on the NNGP priors, the conditional distribution can be computed for all types of reference sets. So, based on the recursive representation we can relax the nested condition on the NNCGP nested reference set. Specifically, 
\begin{align}
 \hat{y}_{t-1}(\bs)|\bfZ_{t-1},\hat{y}_{t-2}(\bs), \bftheta_{t-1},\bfbeta_{t-1} \sim & N(\zeta_{t-2}\hat{y}_{t-2}(\bs)+\mathbf{h}^T_{t-1}(\bs)\bfbeta_{t-1}+V_{t-1,\bs}\mu_{t-1,\bs},V_{t-1,\bs}), \label{response_nncgp_yimputation}%
\end{align}
 with, $\mu_{t-1,\bs}= V_{t-1,\bs}^{-1}\mathbf{B}_{t-1,\bs}\bigl[z_{t-1}(N_{t-1}(\bs))-\mathbf{h}^T_{t-1}(N_{t-1}(\bs))\bfbeta_{t-1}-\zeta_{t-2}(N_{t-1}(\bs))\circ \hat{y}_{t-2}(N_{t-1}(\bs))\bigr]$, $\mathbf{B}_{t,\bs} = C_{\bs,N_t(\bs)}^TC_{N_t{\bs}}^{-1}$, and $V_{t,\bs} = C(\bs,\bs)-C_{\bs,N_t(\bs)}^TC_{N_t{\bs}}^{-1}C_{\bs,N_t(\bs)}$. The $\circ$ represents the Hadamard product between two matrices.

Using the Markovian property of the co-kriging model \citep{o1998markov}, the joint likelihood of the proposed model in (\ref{eq:recursive}) can be factorized as a product of likelihoods at different fidelity levels conditional on  $\hat{y}_{t-1}(\bfS_{t})=\{\hat{y}_{t-1}(\bs_{t,1}),\dots,\hat{y}_{t-1}(\bs_{t,n_{t}})\}$ for $t=2,\dots,T$ and prior $\bfw_{t}$ for $t=1,...,T$,  i.e.: 
\begin{align}
L(\boldsymbol{Z}_{1:T}|\cdot) & =p(\bfZ_{1}|\bfw_{1},\bfbeta_{1},\tau_{1})\prod_{t=2}^{T}p(\bfZ_{t}|\bfw_{t},\bfbeta_{t},\hat{y}_{t-1}(\bfS_{t}),\bfgamma_{t-1},\tau_{t})\nonumber \\
 & =N(\bfZ_{1}|\mathbf{h}_{1}(\bfS_{1})\bfbeta_{1}+\bfw_{1},\tau_{1}\mathbf{I})\prod_{t=2}^{T}N(\bfZ_{t}|\zeta_{t-1}(\bfS_{t})\circ \hat{y}_{t-1}(\bfS_{t})+\mathbf{h}_{t}(\bfS_{t})\bfbeta_{t}+\bfw_{t},\tau_{t}\mathbf{I}),\label{conditional_ind}
\end{align}
where $(\cdot)$ denotes all the parameters associated with the model. This representation makes it possible to integrate out the latent variable $\bfw_t$ independently for each fidelity level  $t=1,\dots, T$ .  

\subsection{Collapsed Recursive Nearest Neighbor Co-kriging Model}

We represent the multivariate Gaussian latent variable $\bfw_{t}(\bfS_{t})$ as a linear model:
\begin{align*}
    w_{t}(\bs_{t,1}) & =0+\eta_{t,1}, \\
    w_{t}(\bs_{t,i}) & =a_{t,i,1}w_{t}(\bs_{t,1})+a_{t,i,2}w_{t}(\bs_{t,2})+\dots+a_{t,i,i-1}w_{t}(\bs_{t,i-1})+\eta_{t,i}, \text{for $i=2,\dots, n_t$} 
\end{align*}
for $t=1,\dots,T$. We set $\eta_{t,i} \sim N(0,d_{t,i,i})$ independently for all $t,i$, $d_{t,1,1}= \var(w_{t,1})$ and $d_{t,i,i}= \var(w_{t,i}|\{w_{t,j}; j<i \})$ for $i=2,\dots,n_t$ and  $t=1,\dots,T$. In a matrix form we can write $\bfw_{t}(\bfS_{t}) =\bfA_{t}\bfw_{t}(\bfS_{t}) +\bfeta_{t}$, where  $\bfA_{t}$ is an $n\times n$ strictly lower-triangular matrix and $\bfeta_{t}\sim N(0,\bfD)$ and  $\bfD$ is diagonal. 
Based on the structure of $\bfA_t$,  we can write the covariance of each level as $\bfC_t(\bftheta_t) = (\mathbf{I}_t - \bfA_t)^{-1}\bfD_t(\mathbf{I}_t - \bfA_t)^{-T}$. The NNGP prior constructs a sparse strictly lower triangular matrix $\bfA$ with no more than $m \ (\text{where } m \ll n)$ non-zero entries in each row resulting in an approximation of the covariance matrix $\bfC_t$. So the approximated inverse $\tilde{\bfC}_t^{-1}(\bftheta_t)= (\mathbf{I}_t - \bfA_t)\bfD_t^{-1}(\mathbf{I}_t - \bfA_t)^{T}$ is a sparse matrix and can be computed based on $\mathcal{O}(n_t m^3)$ operations. 

We call the integrated version of the above model  collapsed RNNC model.  Specifically, after integrating out ${\bf w}_t$ the proposed RNNC model can be written as: 
\begin{align}
z_1(\mathbf{S}_1)|\bfbeta_1,\bftheta_1,\tau_1 & \sim N(\mathbf{h}_1^T(\mathbf{S}_1)\bfbeta_1, \tilde{\Lambda}_1(\bfS_1,\bftheta_1,\tau_1)), \nonumber  \\
z_t(\mathbf{S}_t)| \bfbeta_t,\bftheta_t,\tau_t, \zeta_{t}(\bfS_t), \hat{y}_{t-1}(\bfS_t) & \sim N(\zeta_t(\bfS_t)\circ \hat{y}_{t-1}(\bfS_t)+\mathbf{h}_t^T(\mathbf{S}_t)\bfbeta_t,\tilde{\bfLambda}_t), \label{collapse_structure_NNCGP_AR}
\end{align}
for $t=2,\ldots,T$, where $\tilde{\Lambda}_t(\bftheta_t,\tau_t)  = \tilde{C}_t(\bftheta_t)+\tau_t^2\mathbf{I}=\sigma_{t}^2\tilde{\mathbf{R}}_{t}(\bfphi_t)+\tau_t^2\mathbf{I}$ is the covariance matrix of the observations,  $\tilde{C}_t(\bftheta_t)$ is the sparse covariance matrix with parameters $\boldsymbol{\theta}_t=\{\sigma_t^2,\boldsymbol{\phi}_t\}$ and $\tau_t^2$ is the variance of the error $\epsilon_t$ at level $t$.  By applying Sherman-Morrison-Woodbury formula, the inverse and determinant of $\tilde{\boldsymbol{\Lambda}}$ get the computationally convenient form
\begin{align*}
\tilde{\boldsymbol{\Lambda}}^{-1}_t = \tau^{-2}_t\mathbf{I}-\tau^{-4}_t(\tilde{\bfC}_t(\bftheta_t)^{-1}+\tau^{-2}_t\mathbf{I})^{-1},\\
\text{det}(\tilde{\boldsymbol{\Lambda}}_t)=\tau^{2n}_t\text{det}(\tilde{\bfC}_t(\bftheta_t))\text{det}(\tilde{\bfC}_t(\bftheta_t)^{-1}+\tau^{-2}_t\mathbf{I}).
\end{align*} 

For simplicity let us denote $\bfTheta_{t}=(\bfbeta_t,\gamma_t,\bftheta_t,\tau_t)$. Based on this representation, the joint posterior approximation of all the unknowns is:
\begin{align}
\small
    p(\bfTheta_{1:T}, \hat{\bfy}_{1:T-1}(\bfS_{2:T}^*)| \bfZ_{1:T}) & = p(\bfTheta_{1}|\bfZ_{1}) \prod_{t=2}^{T}  p(\bfTheta_{t}, \hat{\bfy}_{1}(\bfS_{t}^*)|\bfZ_{t})    \nonumber  \\ & \propto p(\bfTheta_{1}) \tilde{L}(\bfZ_{1}|\bfTheta_{1}) \prod_{t=2}^{T} p(\bfTheta_{t})\tilde{L}(\bfZ_{t}|\bfTheta_{t},\hat{\bfy}_{t-1}(\bfS_{t}))
  \tilde{p}(\hat{\bfy}_{t-1}(\bfS_{t}^*)|\cdot) , \label{eq:dsvsgsdbsbg}
\end{align}
where $\tilde{L}(\bfZ_{t}|\bfTheta_{t},\hat{\bfy}_{t-1}(\bfS_{t}))$ is the approximated likelihood using the sparse representation and $\tilde{p}(\hat{\bfy}_{t-1}(\bfS_{t}^*)|\cdot)$ the nearest neighbor Gaussian process prediction at locations  $\bfS_{t}^{*}=\bigcup\limits _{i=t+1}^{T}\bfS_{i}\backslash\bfS_{t}=\{s_{t,1}^{*},\ldots,s_{t,n_{t}^{*}}^{*}\}$ as a set of knots of fidelity level $t$. This contains the observed locations that are not in the $t^{th}$ level but in the higher fidelity levels.  The $(\cdot)$ represents the parameters and data necessary to produce the prediction distortion at level $t-1$.  Note that the  prediction probability can be excluded for cases with hierarchically nested  structure for the spatial locations.  For each level, a  Gibbs sampler can be employed to fascilitate inference based on the conditional representation $p(\bfTheta_{T}|\hat{\bfy}_{t-1}(\bfS_{t}),\bfZ_{T})$ and  $p(\hat{\bfy}_{t-1}(\bfS_{t}^*)|\bfTheta_{T},\bfZ_{T})$ which is given in Eq.~(4.2).

For $\bfTheta_{t}=(\bfbeta_t,\gamma_t,\bftheta_t,\tau_t)$, by assigning independent conjugate prior $\bfbeta_t \sim N(\bfmu_{\bfbeta_t},\mathbf{V}_{\bfbeta_t})$ and $\gamma_t \sim N(\bfmu_{\gamma_t},\mathbf{V}_{\gamma_{t}})$,  
we achieve explicit forms of the conditional distribution for parameters $\bfbeta_t$ and $\gamma_{t}$ as:
\begin{align}
&\bfbeta_t|\bfZ_t,\hat{\bfy}_{t-1}(\bfS_{t}),\bftheta_t, \bfgamma_t\tau_t^2 \sim N(\mathbf{V}_{\beta_t}^*\bfmu_{\beta_t}^*,\mathbf{V}_{\beta_t}^*), \label{collapse_NNCGP_beta} \\
&\bfgamma_t|\bfZ_t,\hat{\bfy}_{t-1}(\bfS_{t}),\bftheta_t,\bfbeta_t,\tau_t^2 \sim N(\mathbf{V}_{\gamma_{t}}^*\bfmu_{\gamma_{t}}^*,\mathbf{V}_{\gamma_{t}}^*),\label{collapse_NNCGP_gamma}
\end{align}
where  $\mathbf{V}_{\beta_t}^*\bfmu_{\beta_t}^*, \mathbf{V}_{\gamma_{t}}^*,\bfmu_{\gamma_{t}}^*$ are given in Appendix C.  Finally, for each fidelity level $t$, we use a Metropolis-Hastings (MH) algorithm targeting the distribution $p(\bftheta_t,\tau_t^2|\bfZ_t,\hat{\bfy}_{t-1}(\bfS_{t}),\bfbeta_t)$ to carry
out the inference.

In the special case of hierarchical nested  structure for the spatial locations, we can avoid sampling from $p(\hat{\bfy}_{t-1}(\bfS_{t}^*)|\bfTheta_{T},\bfZ_{T})$ using the observed locations  $ z_{t-1}(\bfS_{t}^*)$.  We can prove that the mean and variance of the predictive distribution at level $T$ of the collapsed RNNC is the same as the mean and variance of the predictive distribution of the NNCGP. The proof is very similar to \citet{le2014recursive} in the sense that we just need to substitute the GP priors with the NNGP priors and add the nugget effect in each level. 

\subsection{Conjugate Recursive  Nearest Neighbor Co-kriging Model}

Both NNCGP and collapsed RNNC models rely on the MCMC inference which can be practically prohibitive when analyzing thousands or millions of spatial data sets. Following recent work by \cite{finley2019JCGS}, we propose a MCMC free procedure to achieve exact Bayesian inference at a more practical time. Because the computational efficiency  of the  estimation procedure in \cite{finley2019JCGS} is sensitive to the number of parameters, it cannot be applied directly to our model.  We utilize the RNNC model  conditionally independent posterior representation to decompose the parametric space into smaller different groups based on the fidelity levels. To make MCMC free inference possible, we  re-parameterize the covariance function of the collapsed recursive co-kriging model  as $\tilde{\bfLambda}_t(\btheta_t,\tilde{\tau}_t^2)= \sigma_t^2 \tilde{\bfSigma}_t$,  where $\tilde{\bfSigma}_t = \tilde{\mathbf{R}}_{t}+\tilde{\tau}_t^2\mathbf{I}$
, $\tilde{\mathbf{R}}_{t}$ is the nearest-neighbor approximation correlation matrix, and $\tilde{\tau}_t^2 = \frac{\tau_t^2}{\sigma_t^2}$.   To avoid the computational bottleneck due to the MCMC, we propose to make  fast estimation $(\bfphi_t,\tilde{\tau}_t^2)$ through a cross-validation approach for each level as well as use the prediction means of $\bfy_{t-1}(\bfS_{t})$ based on the estimated values. We estimate $\hat{y}_{t}(\bfS_t^*)$ by the posterior mean $\bar{\hat{y}}_t(\bfS_t^*) = \mathbf{1}_{t>1}(t)\mathbf{g}_{t-1}^T(\bfS_t^*)\hat{\bfgamma}_{t-1}\hat{y}_{t-1}(\bfS_t^*)+\mathbf{h}_{t}^T(\bfS_t^*)\hat{\bfbeta}_t + V_{t,\bfS_t^*}\mu_{t,\bfS_t^*}$. In the case that we have nested locations,  $y_t(\bs_u)$  for a location $\bs_u \in \bfS_{t-1}$ is estimated with an empirical approach $\bar{\hat{y}}_{t-1}(\bs_u)$ as $z_{t-1}(\bs_u)$ and its variance is equal to the variance of the nugget effect. Given $\bfphi_t,\tilde{\tau}_t^2$ and $\hat{y}_{t}(\bfS_t)$, the convariance matrix $\tilde{\bfSigma}_t$ can be calculated analytically. 

For computational convenience, we assign an independent conjugate prior for the parameters of each level such as $p(\bfbeta_1, \dots,\bfbeta_T,\sigma_1^2, \dots, \sigma_T^2,\bfgamma_1, \dots,\bfgamma_T)=\prod_{t=1}^Tp(\bfbeta_t)p(\sigma_t^2)p(\bfgamma_t)$ such as $\bfbeta_t \sim N(\bfmu_{\bfbeta_t},\sigma_t^2\mathbf{V}_{\bfbeta_t})$, $\sigma_t^2 \sim IG(a_t,b_t)$, and $\bfgamma_t \sim N(\bfmu_{\bfgamma_{t}},\sigma_{t+1}^2\mathbf{V}_{\bfgamma_{t}})$. Based on this specifications, the posterior density function can be separated for each level $t$ such as: 
\begin{align}
    p(\bfbeta_t,\bfgamma_{t-1},\sigma_t^2 &|\bfZ_t,\hat{y}_{t-1}(\bfS_t)) \propto  IG(\sigma_t^2|a_t,b_t)N(\bfbeta_t|\bfmu_{\bfbeta_t},\sigma_t^2\mathbf{V}_{\bfbeta_t})N(\bfgamma_{t-1}|\bfmu_{\bfgamma_{t-1}},\sigma_t^2\mathbf{V}_{\bfgamma_{t-1}}) \nonumber \\
    & \times N(\bfZ_t|\zeta_{t-1}(\bfS_t)\circ \hat{y}_{t-1}(\bfS_t)+\mathbf{h}^T_t\bfbeta_t,\sigma_t^2\tilde{\bfSigma}_t).
\end{align}

We can compute the full conditional density function of $\bfgamma_{t-1},\bfbeta_t$, and $\sigma_t^2$ as 
\begin{align}
\bfgamma_{t-1}|\bfbeta_t,\sigma_t^2,\bfZ_t,\hat{y}_{t-1}(\bfS_t)) & \sim  N(\bfgamma_{t-1}|\tilde{\mathbf{V}}_{\bfgamma_{t-1}}\tilde{\bfmu}_{\bfgamma_{t-1}},\sigma^2\tilde{\mathbf{V}}_{\bfgamma_{t-1}}),  \label{conjugate_gamma}\\
   \bfbeta_{t}|\sigma_t^2,\bfZ_t,\hat{y}_{t-1}(\bfS_t) & \sim N(\bfbeta_t|\tilde{\mathbf{V}}_{\beta_t}\tilde{\bfmu}_{\beta_t}, \sigma_t^2\tilde{\mathbf{V}}_{\beta_t}) \label{conjugate_beta}\\
   \sigma_t^2|\bfZ_t,\hat{y}_{t-1}(\bfS_t) & \sim IG(\sigma_t^2|a_t^*, b_t^*) \label{conjugate_sigma}
\end{align}
were $\tilde{\mathbf{V}}_{\bfgamma_{t-1}},\tilde{\bfmu}_{\bfgamma_{t-1}},\tilde{\mathbf{V}}_{\beta_t}\tilde{\bfmu}_{\beta_t}, \sigma_t^2\tilde{\mathbf{V}}_{\beta_t}, a_t^*$, and  $b_t^*$ are given analytically in Appendix D. Note that for  $t = 1$,   $\bfgamma_{0}$ and $y_{0}(\bfS_t)$ do not exist. Also the conditional posterior density function of $\bfbeta_1$ and $\sigma_1^2$ are slightly different as explained in Appendix D.

\allowdisplaybreaks
\begin{algorithm} 
\begin{description}
\item [{step~1}] Start from fidelity level 1($t=1$), construct a set
$L_{t}$ that contains $l_{t}$ number of candidates of parameters
$\bfphi_{t}$ and $\tilde{\tau}_{t}^{2}$.
\item [{step~2}] Choose a $(\phi_{t},\tilde{\tau_{t}}^{2})$ from $L_{t}$.
Split the data set of fidelity level $t$ into $K$ folds.
\item [{step~3}] Remove $k^{th}$ fold of data set $\bfS_{t}$, denote
as $\bfS_{t,k}$, then estimate $\sigma_{t}^{2}|\bfZ_{t},\hat{y}_{t-1}(\bfS_{t})$
with the posterior mean $\hat{\sigma}_{t}^{2}=\frac{b_{t}^{*}}{a_{t}^{*}-1}$
of \eqref{conjugate_sigma}. Estimate $\bfbeta_{t}|\sigma_{t}^{2},\bfZ_{t},\hat{y}_{t-1}(\bfS_{t})$
with the posterior mean $\hat{\bfbeta}_{t}=\tilde{\mathbf{V}}_{\beta_{t}}\tilde{\bfmu}_{\beta_{t}}$
of \eqref{conjugate_beta}. Estimate $\bfgamma_{t-1}|\bfbeta_{t},\sigma_{t}^{2},\bfZ_{t},\hat{y}_{t-1}(\bfS_{t})$
with the posterior mean $\hat{\bfgamma}_{t-1}=\tilde{\mathbf{V}}_{\gamma_{t-1}}\tilde{\bfmu}_{\gamma_{t-1}}$ of \eqref{conjugate_gamma}.
\item [{step~4}] Predict test data set $z_{t}(\bfS_{t,k})$ by posterior
mean $$\hat{z}_{t}(\bfS_{t,k})=\mathbf{1}_{t>1}(t)\mathbf{g}_{t-1}^{T}(\bfS_{t,k})\hat{\bfgamma}_{t-1}\hat{y}_{t-1}(\bfS_{t,k})+\mathbf{h}_{t}^{T}(\bfS_{t,k})\hat{\bfbeta}_{t}+V_{t,\bfS_{t,k}}\mu_{t,\bfS_{t,k}}.$$
\item [{step~5}] Repeat steps 3-4 over all $K$ folds, calculate the average
root mean square prediction error(RMSPE) by 
\[
\text{RMSPE}=\frac{\sum_{k=1}^{K}\biggl[\sum_{\bs=\bfS_{t,k}}(z_{t}(\bs)-\hat{z}_{t}(\bs))^{2}/n_{k}\biggr]}{K}.
\]
\item [{step~6}] Repeat steps 2-5 over all values in candidate set $L_{t}$,
choose the value of $\hat{\bfphi}_{t}$ and $\hat{\tilde{\tau}}_{t}^{2}$
that minimizes the RMSPE. Repeat step 3 on full data set $\bfS_{t}$
by fixing $\bfphi_{t}=\hat{\bfphi}_{t}$, $\bfsigma_{t}^{2}=\hat{\bfsigma}_{t}^{2}$.
Estimate $\hat{y}_{t}(\bfS_{t}^{*})$ by posterior mean $$\bar{\hat{y}}_{t}(\bfS_{t}^{*})=\mathbf{1}_{t>1}(t)\mathbf{g}_{t-1}^{T}(\bfS_{t}^{*})\hat{\bfgamma}_{t-1}\hat{y}_{t-1}(\bfS_{t}^{*})+\mathbf{h}_{t}^{T}(\bfS_{t}^{*})\hat{\bfbeta}_{t}+V_{t,\bfS_{t}^{*}}\mu_{t,\bfS_{t}^{*}}$$
\item [{step~7}] For a new input location $\bs_{p}$, predict $y_{t}(\bs_{p})$
by posterior: $$\hat{y}_{t-1}(\bs)|\bfZ_{t-1},\hat{y}_{t-2}(\bs))\sim N(\zeta_{t-2}\hat{y}_{t-2}(\bs)+\mathbf{h}_{t-1}^{T}(\bs)\bfbeta_{t-1}+V_{t-1,\bs}\mu_{t-1,\bs},V_{t-1,\bs})$$
Find a confidence interval based on the quantiles of the above distributions.
\item [{step~8}] Repeat steps 1-7 over all $T$ fidelity levels.
\end{description}
{\footnotesize{}\caption{The Algorithm steps for the MCMC free conjugate RNNC procedure. MCMC free
posterior sampling for multi-fidelity level system with $T$ levels.
\label{step:conjugate_nncgp-1}}
}{\footnotesize\par}
\end{algorithm}

A $K$-fold cross-validation method is used for the selection of optimal values for the parameters $\bfphi_t$ and $\tilde{\tau}_t^2$  at level $t$ that provide best prediction performance for the model, from a group of candidates.  The criteria for choosing $\bfphi_t$ and $\tilde{\tau}_t^2$ can be the root mean square prediction error (RMSPE) over the $K$ folds of data set.  The geolocated observations of the $t$ fidelity are partitioned into $K$ equal size subsets. Then, one of the subsets is used as a test set and the others are used for training. The procedure is repeated $K$ times such that each subset is used once as a test set. The computational complexity of these procedures is reduced significantly from the use of the NNGP priors in the recursive co-kriging model. The estimation, tuning and prediction procedure of conjugate RNNC model are given in Algorithm~1. Similar to NNCGP model, the conjugate RNNC model analyzes the data set of each fidelity level sequentially from the lowest level to the highest. For each single fidelity level $t$, the conjugate RNNC model is able to run in parallel for tuning the parameter $\bfphi_t$ and $\tilde{\tau}_t^2$ using a $K$ fold cross validation procedure.  Step $3$, for given values of $(\bfphi_t, \tilde{\tau}_t^2)$, to estimate $(\hat{\bfbeta}_t, \hat{\sigma}_t^2)$ requires $\mathcal{O}(n_tm^3+n_tmp_t^2)$ floating point operations (flops) where $p_t$ is the dimension of $\bfbeta_t$.  Step 6, to  predict at new locations $\bfS_{t}^{*}$ requires $\mathcal{O}(n_t^*m^3)$ flops where $n_t^*$ is the dimension of $\bfS_{t}^{*}$.  Step 7, to  predict at a new location requires $\mathcal{O}(m^3)$ flops.  When we use parallel computing within a fidelity level for parameters   $(\bfphi_t, \tilde{\tau}_t^2)$  the computation in step $3$ becomes extremely fast.  The conjugate RNNC model provides an empirical estimation of spatial effect parameter $\bfphi_t$ and noise parameter $\tilde{\tau_t}^2$ within a given resolution. We note that the proposed MCMC free inference can be viewed as a sequential optimization technique which splits the parametric space into several lower dimension components where we can apply conditional independent conjugate NNGP models.

\section{Synthetic Data Example and Real Data Analysis}

We study the performance of our proposed procedures, the conjugate RNNC and the collapsed RNNC, as well as compare their performances with that of the sequential NNCGP. The empirical study is based on one synthetic data set example with nested and one with non-nested input data sets. Also we use a real satellite data set application.  As measures of performance we use the root mean squared prediction errors (RMSPE), coverage probability of the $95\%$ equal tail credible interval (CVG($95\%$)), average length of the $95\%$ equal tail credible interval (ALCI($95\%$)), and continuous rank probability score  (CRPS) \citep{GneitingRaftery2007}. Details of these measurements are given in Appendix~\ref{Metrics}. The simulations were performed in MATLAB R2018a, on a personal computer with specifications (intelR i7-3770 3.4GHz Processor, RAM 8.00GB, MS Windows 64bit).We have also included a simulation study with four levels of fidelity in the supplementary materials.

\subsection{Simulation Study}
\begin{figure}
\centering
\subfloat[Non-nested low fidelity observations\label{fig:full_gp_nest_prediction_gp}]{\centering \includegraphics[width=0.49\linewidth]{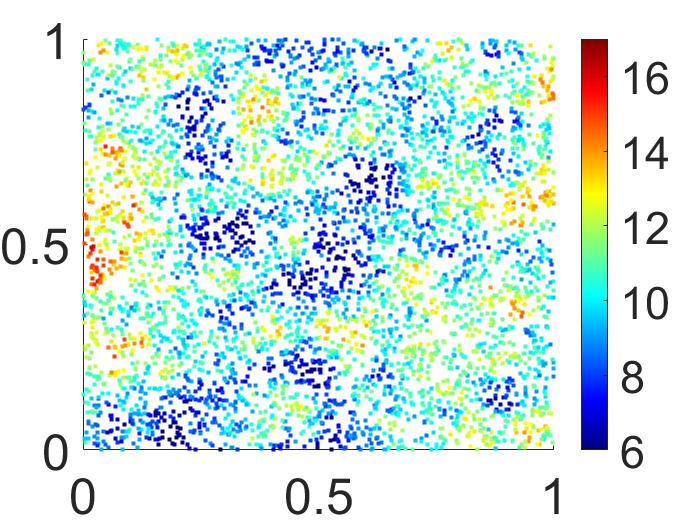}
}
\subfloat[Non-nested high fidelity observations\label{fig:full_gp_nest_prediction_nngp}]{\centering \includegraphics[width=0.49\linewidth]{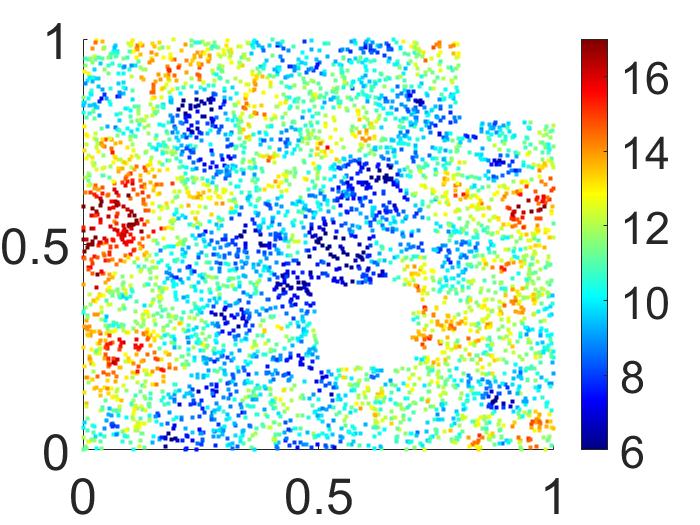}
}
\caption{Observations for two fidelity
level structure of non-nested observed input space. White boxes indicate the testing regions.{\large{}{}\label{fig:nested_and_non-nested_data_chapter3}}}
\end{figure}

We consider a two-fidelity level system represented by the hierarchical
statistical model  (3.1) defined  on a two dimensional unit
square domain with univariate observation data sets for both $\bfZ_{1}$
and $\bfZ_{2}$. Let the design matrix be $\mathbf{h}(\bfS_{t})=\mathbbm{1}$, the 
autoregressive coefficient function be an unknown constant $\zeta_{1}(\bs) = \gamma_1$, and exponential covariance functions. We generate two synthetic data sets for the above statistical model. The true values of the parameters
are listed in Table \ref{tab:Univariate-two-fildelity-nested-chapter3} and Table \ref{tab:Univariate-two-fildelity-nonnested-chapter3}. The data sets on the  nested spatial locations consists of observations $\bfZ_{1}$ and $\bfZ_{2}$
from $100\times100$ grids $\bfS_{1}$ and $\bfS_{2}$, respectively.
The data sets, shown in Figures \ref{fig:full_gp_nest_prediction_gp} and \ref{fig:full_gp_nest_prediction_nngp} are based on a fully
non-nested input where the low fidelity observations $\bfZ_{1}$ and the high fidelity observations $\bfZ_{2}$ are generated at irregularly located at point in sets $\bfS_{1}$ and $\bfS_{2}$ of size
$5000$, while $\bfS_{1}\cap\bfS_{2}=\emptyset$.
In all data sets, a few small square regions from $\bfZ_{2}$ are
treated as a testing data-set, and the rest of $\bfZ_{2}$ and $\bfZ_{1}$
are treated as training data sets. The testing regions for the non-nested input can be seen as white boxes in Figure 2(b).

Regarding the Bayesian inference, we compared the sequential NNCGP model, with the proposed collapsed RNNC model and that with the  conjugate RNNC model, on both nested and non-nested data sets. We assigned similar non-informative priors for all the four models. We assign independent conjugate prior on parameters $\beta_{1}\sim N(0,1000)$, $\beta_2\sim N(0,1000)$, and scale parameter $\gamma_1$. We assign independent inverse gamma prior on spatial variance parameters $\sigma_1^2\sim IG(2,1)$, and $\sigma_2^2\sim IG(2,1)$ and on the noise parameters $\tau_1^2\sim IG(2,1)$, $\tau_2^2 \sim IG(2,1)$. We also assign uniform prior on the range parameters $\phi_1\sim U(0,25)$, and $\phi_2\sim U(0,25)$. For the collapsed RNNC model, we run Markov chain Monte
Carlo (MCMC) samplers for $35000$ iterations where the first $5000$
iterations are discarded as a burn-in, and convergence of the MCMC
sampler was diagnosed from the individual trace plots. The RMSPE with  a 5-fold cross-validation was used for the conjugate RNNC model. We select $(\phi_t,\tilde{\tau}_t^2)$ on a grid, for $\phi_t$ from the range  $[0.1 , 25]$, and for $\tilde{\tau}_t^2$ from the range $[0.0005 , 0.4]$.  No significant differences were observed when we used 3-fold cross-validation and 7-fold cross-validation approach.

\begin{table}[ht]
\centering
{\footnotesize{}{}\centering}%
\begin{tabular}{c|c|cc|cc|c}
\hline 
\multirow{1}{*}{} & \multirow{1}{*}{{\footnotesize{}{}True}} & \multicolumn{5}{c}{{\footnotesize{}{}Nested data-set}}
\tabularnewline
 & {\footnotesize{}{}values}  & \multicolumn{2}{c}{{\footnotesize{}{}Sequential NNCGP}} & \multicolumn{2}{c}{{\footnotesize{}{}Collapsed RNNC}}  & \multicolumn{1}{c}{{\footnotesize{}{}Conjugate RNNC}}\tabularnewline
\hline 
{\footnotesize{}{}$\beta_{1}$}  & {\footnotesize{}{}10}  & {\footnotesize{}{}10.29 }  & {\footnotesize{}{}(9.93,10.57)} &  {\footnotesize{}{}9.96 }  & {\footnotesize{}{}(9.60,10.32)} &   {\footnotesize{}{}10.02}  
\tabularnewline
{\footnotesize{}{}$\beta_{2}$}  & {\footnotesize{}{}1} & {\footnotesize{}{}0.77}  & {\footnotesize{}{}(0.59,1.04)} & {\footnotesize{}{}0.87}  & {\footnotesize{}{}(0.59,1.13)} &  {\footnotesize{}{}0.82}  
\tabularnewline
{\footnotesize{}{}$\sigma_{1}^{2}$}  & {\footnotesize{}{}4}& {\footnotesize{}{}3.55}  & {\footnotesize{}{}(2.77,4.38)} & {\footnotesize{}{}3.46}  & {\footnotesize{}{}(2.96,4.27)} &  {\footnotesize{}{}3.15}  
\tabularnewline
{\footnotesize{}{}$\sigma_{2}^{2}$}  & {\footnotesize{}{}1}& {\footnotesize{}{}0.81}  & {\footnotesize{}{}(0.27, 2.05)} & {\footnotesize{}{}0.98}  & {\footnotesize{}{}(0.43, 1.88)} &  {\footnotesize{}{}0.79} 
\tabularnewline
{\footnotesize{}{}$1/\phi_{1}$}  & {\footnotesize{}{}10} & {\footnotesize{}{}10.42}  & {\footnotesize{}{}(8.15,13.47)} & {\footnotesize{}{}10.50}  & {\footnotesize{}{}(8.59,13.90)} &  {\footnotesize{}{}12.1}
\tabularnewline
{\footnotesize{}{}$1/\phi_{2}$}  & {\footnotesize{}{}10} & {\footnotesize{}{}14.96}  & {\footnotesize{}{}(3.37, 20.29)} & {\footnotesize{}{}15.69}  & {\footnotesize{}{}(5.92, 19.98)} &  {\footnotesize{}{}19.6}  
\tabularnewline
{\footnotesize{}{}$\gamma_{1}$}  & {\footnotesize{}{}1} & {\footnotesize{}{}0.99}  & {\footnotesize{}{}(0.98,1.00)} & {\footnotesize{}{}0.99}  & {\footnotesize{}{}(0.98,1.00)} &  {\footnotesize{}{}0.99}    \tabularnewline
{\footnotesize{}{}$\tau_{1}^{2}$}  & {\footnotesize{}{}0.1}  & {\footnotesize{}{}0.12}  & {\footnotesize{}{}(0.10,0.14)} & {\footnotesize{}{}0.15}  & {\footnotesize{}{}(0.10,0.19)} &  {\footnotesize{}{}0.12}      \tabularnewline
{\footnotesize{}{}$\tau_{2}^{2}$}  & {\footnotesize{}{}0.05} & {\footnotesize{}{}0.07}  & {\footnotesize{}{}(0.03,0.11)} & {\footnotesize{}{}0.10}  & {\footnotesize{}{}(0.04,0.19)} &  \footnotesize{}{}0.16{\footnotesize{} }  \tabularnewline
{\footnotesize{}{}$m$}  & {\footnotesize{}{}10}  & {\footnotesize{}{}-}  & {\footnotesize{}{}-}  & {\footnotesize{}{}-}  & -   & {\footnotesize{}{}-}   \tabularnewline
\hline 
\end{tabular}{\footnotesize{}{}\caption{The estimation of parameters in nested input dataset, using sequential NNCGP, collapsed RNNC and conjugate RNNC models.  \label{tab:Univariate-two-fildelity-nested-chapter3}}
} 
\end{table}

\begin{table}[ht]
\centering
\begin{tabular}{c|ccc}
\hline 
\multirow{1}{*}{} & \multicolumn{3}{c}{{\footnotesize{}{}Nested data-set}} \tabularnewline
 & {\footnotesize{}{}Sequential NNCGP}  & {\footnotesize{}{}Collapsed RNNC}   & {\footnotesize{}{}Conjugate RNNC}  \tabularnewline
\hline 
{\footnotesize{}{}RMSPE} & {\footnotesize{}{}0.63} & {\footnotesize{}{}0.69}&  {\footnotesize{}{}0.72}   \tabularnewline
{\footnotesize{}{}NSME } & {\footnotesize{}{}0.77}  & {\footnotesize{}{}0.75 }&  {\footnotesize{}{}0.71}
\tabularnewline
{\footnotesize{}{}CRPS} & {\footnotesize{}{}0.45}  & {\footnotesize{}{}0.44 }&  {\footnotesize{}{}0.41}
\tabularnewline
{\footnotesize{}{}CVG(95\%) } & {\footnotesize{}{}0.91} &  {\footnotesize{}{}0.88 } & {\footnotesize{}{}0.93}
\tabularnewline
{\footnotesize{}{}ALCI(95\%)} & {\footnotesize{}{}1.93}  & {\footnotesize{}{}1.92 }& {\footnotesize{}{}2.54}
\tabularnewline
\hline 
{\footnotesize{}{}Time(Hour) } & {\footnotesize{}{}4.5}  & {\footnotesize{}{}4.7 }&  {\footnotesize{}{}0.08}  \tabularnewline
\hline 
\end{tabular}{\footnotesize{}{} \caption{Performance measures for the predictive ability of the sequential NNCGP model, collapsed RNNC model and conjugate RNNC
model. \label{tab:efficient_nested_metrics_nested_chapter3}}
} 
\end{table}

In Tables \ref{tab:Univariate-two-fildelity-nested-chapter3} and \ref{tab:Univariate-two-fildelity-nonnested-chapter3}, we report the Monte
Carlo estimates of the posterior means and the associated $95\%$
marginal credible intervals of the unknown parameters using the two different NNCGP based procedures: sequential NNCGP, collapsed RNNC, along with the posterior mean and tuned values of parameters using conjugate RNNC, with $m=10$. There is no significant difference in the estimation of parameters for all MCMC based models (NNCGP and collapsed RNNC) and the true values of the parameters are successfully included in the $95\%$ marginal credible intervals. The introduction of latent interpolants may have caused a small overestimation of $\tau_{2}^2$ for all models.  Instead, the  conjugate RNNC is underestimating the variance of the nugget for the second fidelity level.  The parameter estimations can be improved with a semi-nested or nested structure between the observed locations of the fidelity levels, and it is also shown for the auto-regressive co-kriging model in \citet{konomikaragiannisABTCK2019}. The conjugate RNNC model has similar performance on estimating the mean of the parameters  compared to the NNCGP and RNNC models. However, it does not provide uncertainties regarding these estimations.%

\begin{table}
\centering
{\footnotesize{}{}\centering}%
\begin{tabular}{c|c|cc|cc|c}
\hline 
\multirow{1}{*}{} & \multirow{1}{*}{{\footnotesize{}{}True}} & \multicolumn{5}{c}{{\footnotesize{}{} Non-nested data-set}}
\tabularnewline
 & {\footnotesize{}{}values}  & \multicolumn{2}{c}{{\footnotesize{}{}Sequential NNCGP}} & \multicolumn{2}{c}{{\footnotesize{}{}Collapsed RNNC}} &  \multicolumn{1}{c}{{\footnotesize{}{}Conjugate RNNC}}\tabularnewline
\hline 
{\footnotesize{}{}$\beta_{1}$}  & {\footnotesize{}{}10}  & {\footnotesize{}{}9.71}  & {\footnotesize{}{}(9.36, 10.16)} &  {\footnotesize{}{}9.97 }  & {\footnotesize{}{}(9.52,10.41)} &   {\footnotesize{}{}9.71}    \tabularnewline
{\footnotesize{}{}$\beta_{2}$}  & {\footnotesize{}{}1} & {\footnotesize{}{}0.87}  & {\footnotesize{}{}(0.39,1.36)}  & {\footnotesize{}{}1.23}  & {\footnotesize{}{}(0.24,2.19)} &  {\footnotesize{}{}1.27}    \tabularnewline
{\footnotesize{}{}$\sigma_{1}^{2}$}  & {\footnotesize{}{}4}& {\footnotesize{}{}3.51}  & {\footnotesize{}{}(2.71,4.52)}  & {\footnotesize{}{}3.28}  & {\footnotesize{}{}(3.02,3.72)} &  {\footnotesize{}{}3.84}     \tabularnewline
{\footnotesize{}{}$\sigma_{2}^{2}$}  & {\footnotesize{}{}1}& {\footnotesize{}{}1.05}  & {\footnotesize{}{}(0.18,2.31)}  & {\footnotesize{}{}1.00}  & {\footnotesize{}{}(0.64, 1.49)} &  {\footnotesize{}{}1.19}      \tabularnewline
{\footnotesize{}{}$1/\phi_{1}$}  & {\footnotesize{}{}10} & {\footnotesize{}{}10.77}  & {\footnotesize{}{}(8.07,13.91)}   & {\footnotesize{}{}13.23}  & {\footnotesize{}{}(9.93,15.85)} &  {\footnotesize{}{}6.50}     \tabularnewline
{\footnotesize{}{}$1/\phi_{2}$}  & {\footnotesize{}{}10} & {\footnotesize{}{}12.61}  & {\footnotesize{}{}(3.93,24.07)}  & {\footnotesize{}{}16.01}  & {\footnotesize{}{}(12.84, 19.92)} &  {\footnotesize{}{}20.00}      \tabularnewline
{\footnotesize{}{}$\gamma_{1}$}  & {\footnotesize{}{}1} & {\footnotesize{}{}0.99}  & {\footnotesize{}{}(0.98,1.05)}  & {\footnotesize{}{}0.97}  & {\footnotesize{}{}(0.94,0.99)} &  {\footnotesize{}{}0.96}    \tabularnewline
{\footnotesize{}{}$\tau_{1}^{2}$}  & {\footnotesize{}{}0.1}  & {\footnotesize{}{}0.13}  & {\footnotesize{}{}(0.10,0.15)}   & {\footnotesize{}{}0.10}  & {\footnotesize{}{}(0.07,0.15)} &  {\footnotesize{}{}0.11}      \tabularnewline
{\footnotesize{}{}$\tau_{2}^{2}$}  & {\footnotesize{}{}0.05} & {\footnotesize{}{}0.16} & \footnotesize{}{}(0.04,0.23) & {\footnotesize{}{}0.18}  & {\footnotesize{}{}(0.03,0.29)} &  \footnotesize{}{}0.10{\footnotesize{} }    \tabularnewline
{\footnotesize{}{}$m$}  & {\footnotesize{}{}10}  & {\footnotesize{}{}-}  & {\footnotesize{}{}-}  & {\footnotesize{}{}-}  & -   & {\footnotesize{}{}-}
\tabularnewline
\hline 
\end{tabular}{\footnotesize{}{}\caption{The estimation of parameters in non-nested input dataset  using sequential NNCGP, collapsed RNNC and conjugate RNNC models.  \label{tab:Univariate-two-fildelity-nonnested-chapter3}}
} 
\end{table}

\begin{table}
\center%
\begin{tabular}{c|ccc}
\hline 
\multirow{1}{*}{} & \multicolumn{3}{c}{{\footnotesize{}{}Non-nested data-set}} \tabularnewline
 & {\footnotesize{}{}Sequential NNCGP}  & {\footnotesize{}{}Collapsed RNNC}  &  {\footnotesize{}{}Conjugate RNNC}   \tabularnewline
\hline 
{\footnotesize{}{}RMSPE} & {\footnotesize{}{}0.93} & {\footnotesize{}{}0.92}&  {\footnotesize{}{}1.07}   \tabularnewline
{\footnotesize{}{}NSME } & {\footnotesize{}{}0.75}  & {\footnotesize{}{}0.78 }&  {\footnotesize{}{}0.75} 
\tabularnewline
{\footnotesize{}{}CRPS} & {\footnotesize{}{}0.61}  & {\footnotesize{}{}0.61}&  {\footnotesize{}{}0.65}
\tabularnewline
{\footnotesize{}{}CVG(95\%) } & {\footnotesize{}{}0.93} & {\footnotesize{}{}0.97 }&  {\footnotesize{}{} 0.95}
\tabularnewline
{\footnotesize{}{}ALCI(95\%)} & {\footnotesize{}{}1.49}  & {\footnotesize{}{}1.76 }&  {\footnotesize{}{}2.49}
\tabularnewline
\hline 
{\footnotesize{}{}Time(Hour) } & {\footnotesize{}{}4.1} & {\footnotesize{}{}4.3 } & {\footnotesize{}{}0.05}  \tabularnewline
\hline 
\end{tabular}{\footnotesize{}{} \caption{Performance measures for the predictive ability of the Sequential NNCGP model, collapsed RNNC model and conjugate RNNC
model, in non-nested input. \label{tab:efficient_nonnested_metrics_nonnested_chapter3}}
} 
\end{table}

In Table \ref{tab:efficient_nested_metrics_nested_chapter3} and Table \ref{tab:efficient_nonnested_metrics_nonnested_chapter3}, we report standard
performance measures (defined in Appendix~\ref{Metrics}) for the sequential NNCGP, collapsed RNNC and conjugate RNNC with $m=10$ number of neighbours. All performance measures indicate that the collapsed RNNC model has similar predictive ability with the sequential NNCGP model. The conjugate RNNC model produced RMSPE value that is 10\% larger than other NNCGP and collapsed RNNC models, but it is still significantly smaller than the RMSPE values from single level NNGP and combined NNGP. The tables also show that the running time of the collapsed RNNC model is not different from the sequential NNCGP model, this is consistent with our previous discussion since the two procedures have the same computational complexity. We observe that the conjugate RNNC model has extremely smaller running time compared to sequential NNCGP models, since the inference for conjugate RNNC model requires the same amount of running time as one iteration in collapsed RNNC model. It is worth pointing out that the cross validation process and tuning process in Algorithm~1 are independent from each other, which makes conjugate RNNC model benefit from parallel computation environments and greatly reduce computational time.

\begin{figure}
\centering
\subfloat[High-level testing data\label{fig:large_nonnest_prediction_test3}]{\centering\includegraphics[width=0.33\linewidth]{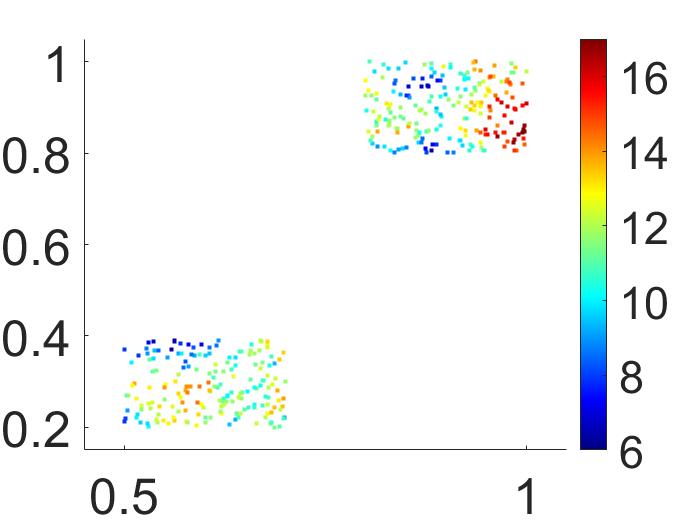}}
\subfloat[\label{fig:large_nonnest_prediction_nncgp3a}Sequential NNCGP prediction]{\centering \includegraphics[width=0.33\linewidth]{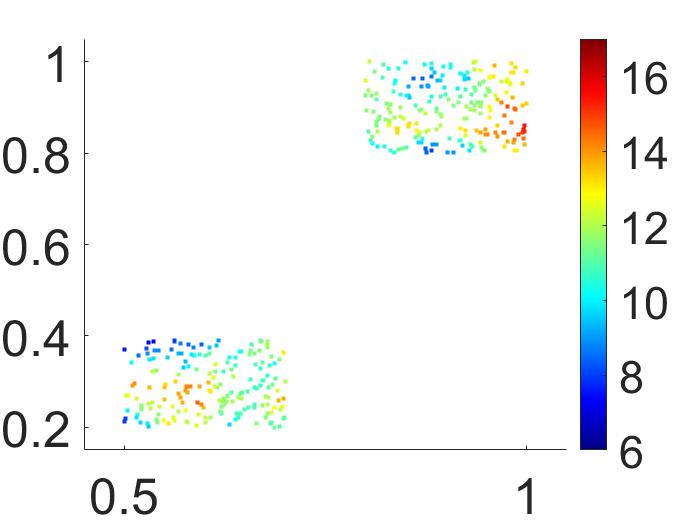}
}
\subfloat[\label{fig:large_nonnest_prediction_nncgp3b}Collapsed RNNC prediction]{\centering \includegraphics[width=0.33\linewidth]{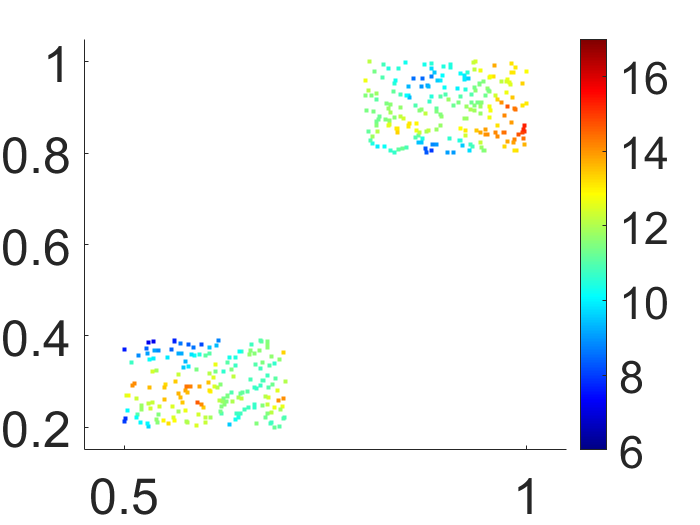}
}
\hfill
\subfloat[Conjugate RNNC prediction\label{fig:large_nonnest_prediction_nngp3}]{\centering \includegraphics[width=0.33\linewidth]{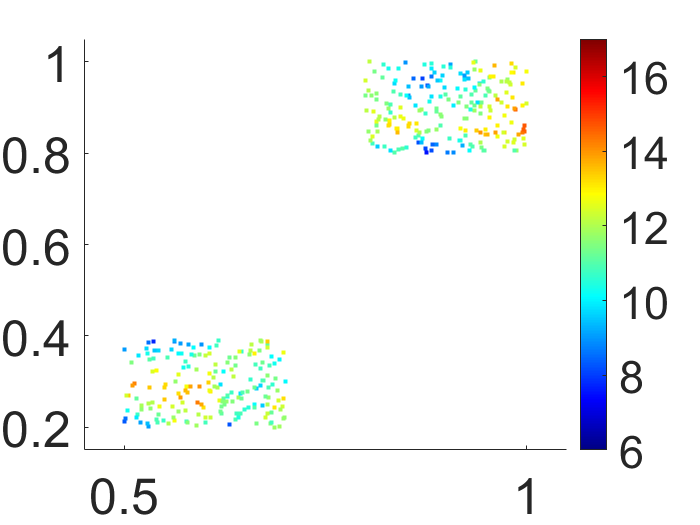}
}
\subfloat[\label{fig:large_nonnest_prediction_nncgp3c}Combined NNGP prediction]{\centering \includegraphics[width=0.33\linewidth]{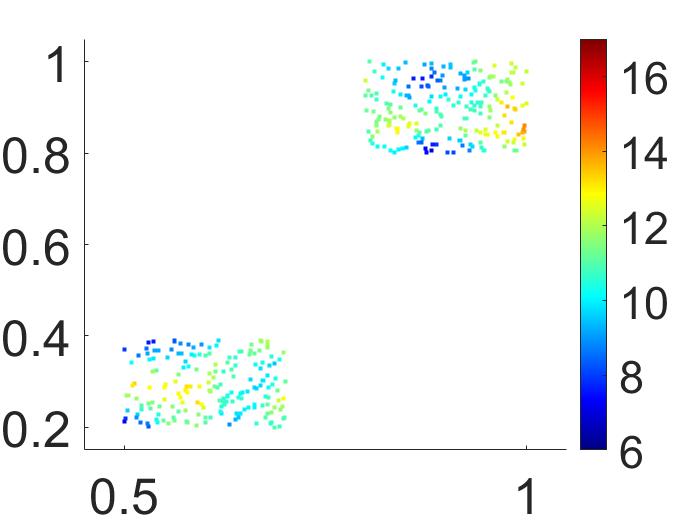}
}
\subfloat[Single level NNGP prediction\label{fig:large_nonnest_prediction_nngp3d}]{\centering \includegraphics[width=0.33\linewidth]{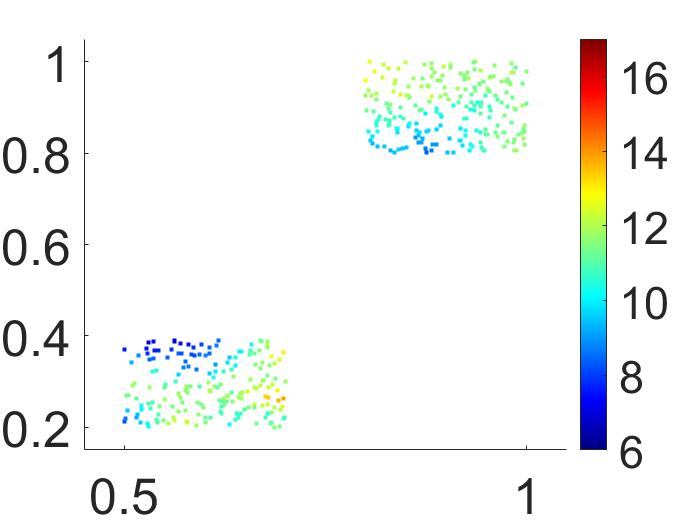}
}
\caption{Non-nested input observations with two fidelity
level structure. Original testing data (a) along with predictions of the high fidelity
level data-set by (b) Sequential NNCGP, (c) Collapsed RNNC, (d) Conjugate RNNC, (e) Combined NNCGP and (f) single level NNCGP.{\large{}{}\label{fig:large_nonnest_prediction_conjugate_plots}}}
\end{figure}

Figure \ref{fig:nested_and_non-nested_data_chapter3} and \ref{fig:large_nonnest_prediction_conjugate_plots} provide the  non-nested synthetic observations
and the prediction plots from sequential NNCGP, collapsed RNNC, conjugate RNNC, combined NNGP and single level NNGP models. We observed that for the testing regions the NNCGP models provides similar prediction surfaces and all NNCGP models has the better presentation of patterns in prediction surface comparing to single level NNGP model.

\subsection{Application to High-resolution Infrared Radiation Sounder data}

We model our data based on the two-fidelity level conjugate RNNC model and on the two-fidelity level sequential NNCGP model. Moreover, we provide comparisons with the single level NNGP model and combined NNGP model.  We consider a linear model for the mean  of the Gaussian processes, in $y_1(\cdot)$ and $\delta_{2}(\cdot)$,  with  linear basis function representation  $\mathbf{h}(s_{t})$
and coefficients $\boldsymbol{\beta}_{t}=\{\beta_{0,t},\beta_{1,t},\beta_{2,t}\}^{T}$. We consider the scalar discrepancy $\zeta(\bs)$ to be unknown constant  and equal to $\gamma$. The number of nearest neighbors $m$ is set to 10, and the spatial process $\bfw_{t}$ is considered  to have a diagonal anisotropic exponential covariance function.

\begin{table}[H]
\begin{adjustbox}{width=1\textwidth}
{}{}\centering %
\begin{tabular}{c|ccccc}
\hline 
\multirow{1}{*}{{}{} } & \multicolumn{5}{c}{{}{}Model}\tabularnewline
 & {}{}Sequential NNCGP  & {}{}Single level NNGP & {}{}Combined NNGP & {}{}Collapsed RNNC & {}{}Conjugate RNNC \tabularnewline
\hline 
{}{}RMSPE  & {}{}1.20  & {}{}1.82& {}{}1.68 & {}{}1.21 & {}{}1.36 \tabularnewline
{}{}NSME  & {}{}0.84  & {}{}0.55 &0.67 & {}{}0.85 & {}{}0.82 \tabularnewline
{}{}CRPS  & {}{}0.70  & {}{}1.65 &0.93 & {}{}0.68 & {}{}0.75 \tabularnewline
{}{}CVG(95\%)  & {}{}0.93  & {}{}0.84 &0.92 & {}{}0.94 & {}{} 0.94 
\tabularnewline
{}{}ALCI(95\%)  & {}{}3.09  & {}{}4.21 &5.79 & {}{}3.16 & {}{} 4.39
\tabularnewline
\hline 
{}{}Time(Hour)  & {}{}38  & {}{}20 & 32 & {}{}40 & {}{}0.3 \tabularnewline
\hline 
\end{tabular}
\end{adjustbox}
\caption{{\large{}{}{}{}{}{}{}\label{real_data_table_conjugate}}Performance measures for the predictive ability of sequential NNCGP, single level NNGP, combined NNGP, collapsed RNNC and conjugate RNNC models in NOAA 14 and NOAA
15 HIRS instrument data analysis.}
\end{table}

We assign independent normal distribution priors with zero mean and large  variances for $\beta_{0,t},\beta_{1,t},\beta_{2,t}$ and 
$\gamma$.  We assign independent uniform prior distributions $U(0,1000)$ to the range correlation parameters $({\phi}_{t,1},{\phi}_{t,2})$ for $t=1,2$. Also, we assign independent $IG(2,1)$ prior distributions for the variance parameters $\sigma_{t}^{2}$ and $\tau_{t}^{2}$. For the Bayesian inference of the sequential NNCGP, we run the MCMC sampler with  of $35,000$ iterations where the first $5,000$ iterations are discarded as a burn-in. For the Bayesian inference of conjugate RNNC, we consider using posterior means as the estimated values for parameters $\beta_{0,t},\beta_{1,t},\beta_{2,t}$, $\sigma_t^2$ and 
$\gamma$; we also use posterior means as the  imputation values for latent process $\tilde{\bfy}_t$ and for the prediction values of $z(s_p)$ at location $s_p\not\in\tilde{\bfS}_t$.

\begin{figure}
\subfloat[NOAA-15 testing data-set\label{fig:HIRS_conjugate_test}]{\centering \includegraphics[width=0.45\linewidth]{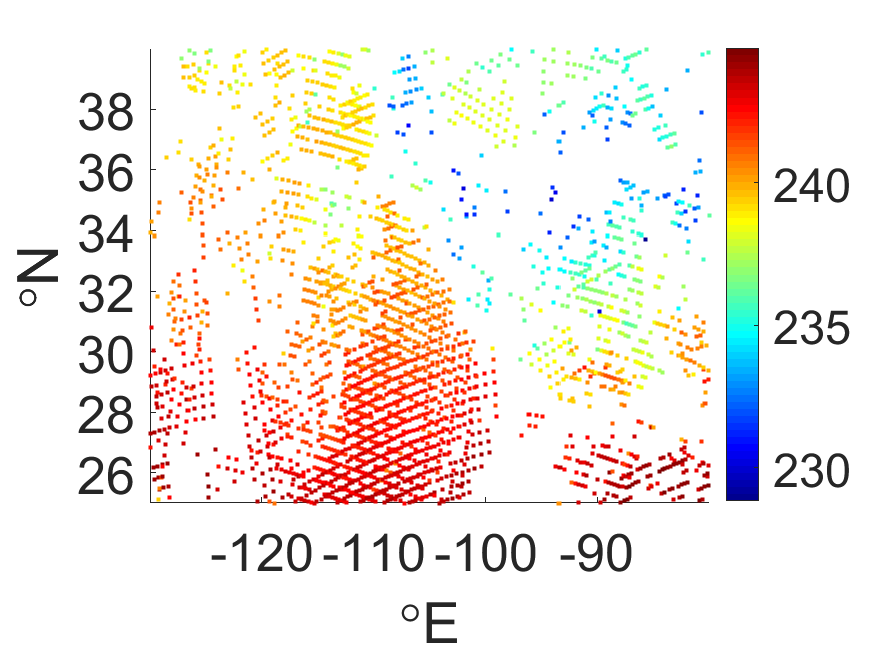}}
\subfloat[\label{fig:HIRS_conjugate_sequentiala}Prediction means by Sequential NNCGP model]{\centering \includegraphics[width=0.45\linewidth]{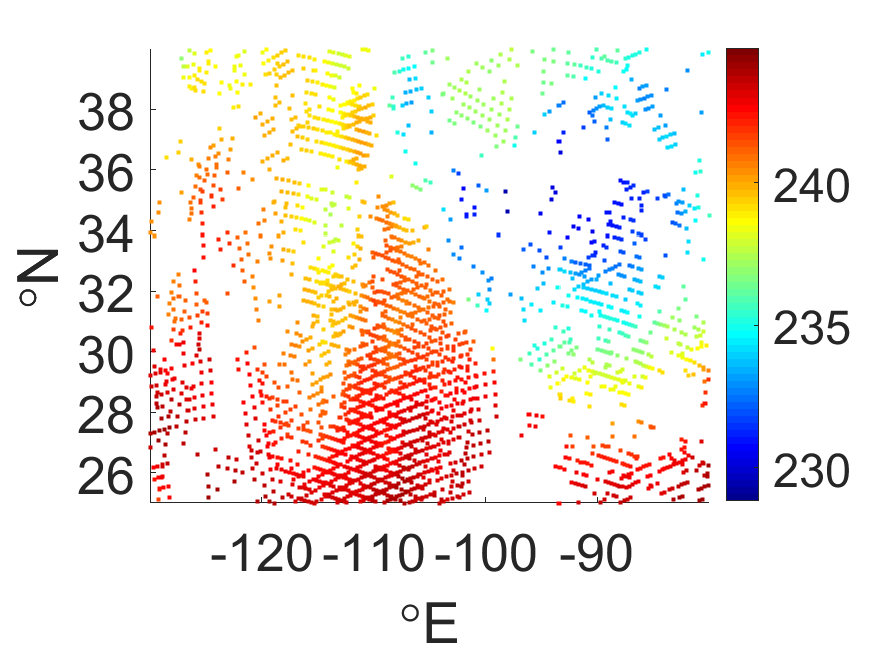}
}
\hfill
\subfloat[\label{fig:HIRS_conjugate_sequentialb}Prediction means by Collapsed RNNC model]{\centering \includegraphics[width=0.45\linewidth]{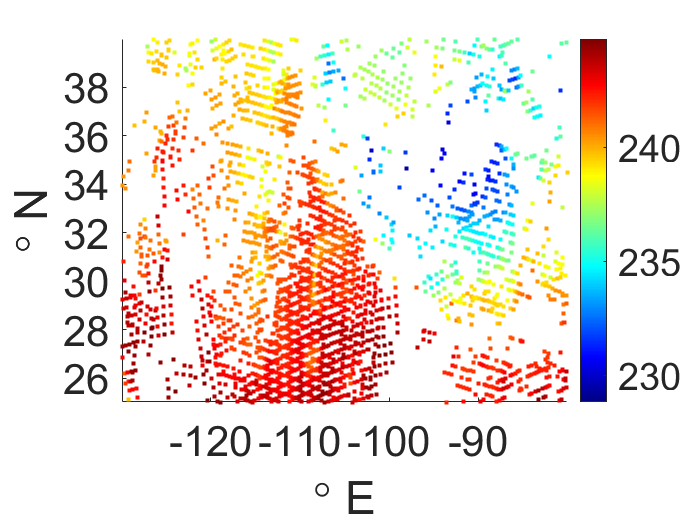}
}
\subfloat[\label{fig:HIRS_conjugate_conjugate} Prediction means by conjugate RNNC
model]{\centering \includegraphics[width=0.45\linewidth]{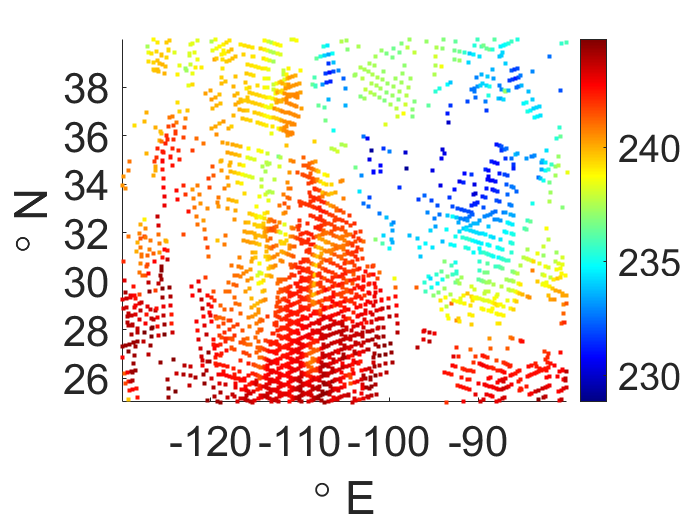}
}
\hfill
\subfloat[\label{fig:HIRS_conjugate_single}Prediction means by single level NNGP
model]{\centering \includegraphics[width=0.45\linewidth]{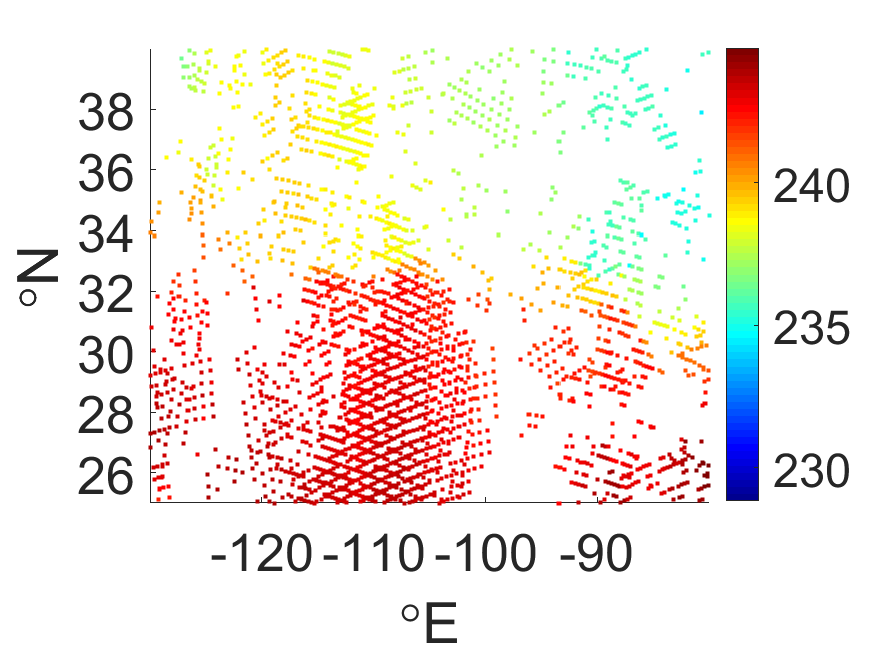}
}
\subfloat[\label{fig:HIRS_conjugate_combine}Prediction means by combined NNGP
model]{\centering \includegraphics[width=0.45\linewidth]{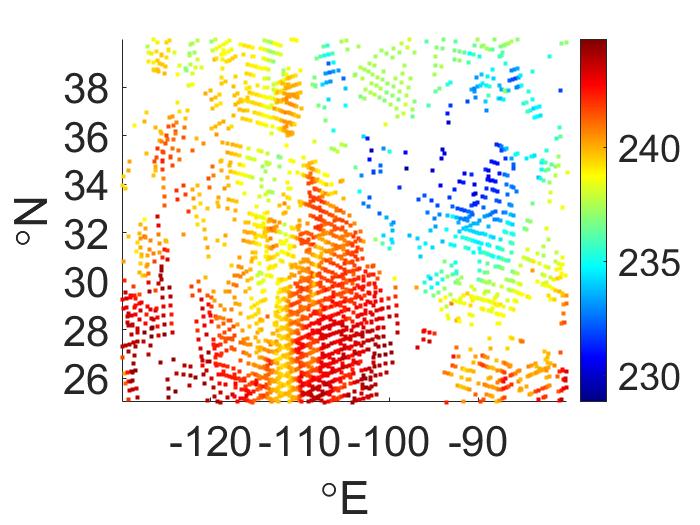}
}
\caption{Predictions of NOAA-15
Brightness Temperatures(K) testing data-set by (b) sequential NNCGP, (c) collapsed RNNC, (d) conjugate
RNNC, (e) single level NNGP and (f) combined NNGP models.\label{fig:Hirs_conjugate_model} }
\end{figure}

The prediction performance metrics of the four different methods are
given in Table \ref{real_data_table_conjugate}. Compared to the single level NNGP model and combined NNGP model, the sequential NNCGP model
 and conjugate RNNC model produced a 20-30\% smaller RMSPE and their NSME is closer to 1. The sequential NNCGP model and collapsed RNNC model also produced larger CVG and smaller ALCI than the single level NNGP model and combined NNGP model. The result suggests that the NNCGP and RNNC models have a substantial improvement in terms of predictive accuracy in real data analysis too. In the prediction plots (Figure \ref{fig:Hirs_conjugate_model}) of the testing data of NOAA-15, we  observe that RNNC models are more capable of capturing the pattern of the testing data than single level NNGP model and combined NNGP model. This is reasonable because the observations from NOAA-14 have provided information of the testing region, and comparing to combined NNGP model, the NNCGP and RNNC models are capable of modeling the discrepancy of observations from different satellites. In the  non-nested structure, the computational complexity of the single level NNGP model is $\mathcal{O}(n_{2}m^{3})$ and that of NNCGP model is $\mathcal{O}((n_{1}+n_{2})m^{3})$, for an MCMC iteration. However, the whole computational complexity of the conjugate RNNC model is $\mathcal{O}((n_{1}+n_{2})m^{3})$ with parallel computational environment, which makes it remarkably computationally efficient without losing significant prediction accuracy. This is consistent with the running times of the models shown in Table \ref{real_data_table_conjugate}.

\begin{figure}
{}{}\centering \includegraphics[width=0.9\linewidth]{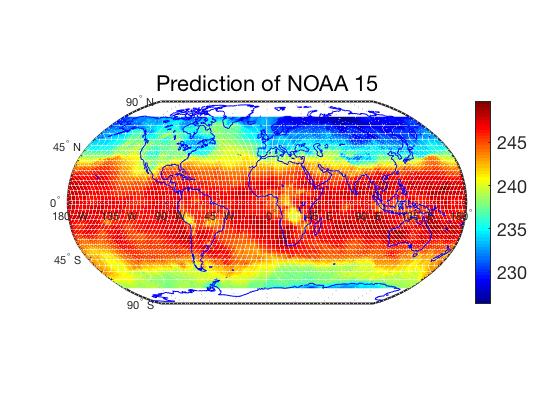}
\caption{{\large{}{}{}{}{}{}{}\label{fig:sfig2-2-1}}The global prediction brightness temperature values  of NOAA 15 using the MCMC free conjugate model.}
\label{fig:fig-7} 
\end{figure}
We apply the MCMC free conjugate RNNC model for gap-filling predictions based upon a discrete global grid. We chose to use $1^{\circ}$ latitude by $1.25^{\circ}$ longitude ($1^{\circ}\times1.25^{\circ}$) pixels as grids with global spatial coverage from $-70^{\circ}$ to $70^{\circ}$N. By applying the NNCGP model, we predict gridded NOAA-15 brightness temperature data on the center of the grids, based on the NOAA-14 and NOAA-15 swath-based spatial support. The prediction plot (Figure \ref{fig:fig-7}) illustrates the ability of the  MCMC free conjugate RNNC model to handle large irregularly spaced data sets and produce a gap-filled composite gridded dataset. The resulting global image of the brightness temperature is practically the same as the sequential NNCGP.

\section{Summary and conclusions}

We  have proposed a new computationally efficient co-kriging method, the recursive nearest neighbor Autoregressive Co-Kriging (RNNC) model, for the analysis of large and multi-fidelity spatial data sets. In particular, we proposed two computationally efficient inferential procedures: a) the  collapsed  RNNC,  and  b) the  conjugate RNNC. Regarding the collapsed RNNC, we integrate out the latent variables  of the RNNC model which  enables the factorisation of the likelihood into terms involving smaller and sparse covariance matrices within each level. Then, a prediction focused approximation is applied to the aforesaid model to further speed up the computation. The cross-validation using grid search on a two or three dimensional space is a computationally feasible method to estimate the hyperparameters. Regarding the proposed conjugate RNNC, it is MCMC free and at most computationally linear in the total number of all spatial locations of all fidelity levels. We compared the proposed collapsed RNNC and conjugate RNNC with  NNCGP in a simulation study and a real data application of intersatellite
calibration. We observed that similar to  NNCGP, the collapsed and conjugate RNNC were also able to improve the accuracy of the prediction
for the HIRS brightness temperatures from the NOAA-15 polar-orbiting
satellite by incorporating information from an older version of the
same HIRS sensor on board the polar orbiting satellite NOAA-14.  The RNNC can be viewed as a  modularization approach to NNCGP model in Bayesian statistics \cite{Bayarri_module} where the analysis is done in steps rather than  jointly.

The proposed procedures can be used for a variety of large multi-fidelity data sets in remote sensing with overlapping areas of observed locations. A natural extension of our model can be done based on a recently proposed sparse plus Low-rank Gaussian Process (SPLGP) \citep{Shirota_Finley2022} who used a combination of Gaussian predictive process  and NNGP in an MCMC-free framework. A natural choice for introducing the non-stationarity in the conjugate RNNC is to use non-dynamic partition methods such as \citep{Heaton_Tech2017, Konomi_env2019}. Moreover, we can use more complex Vechias approximations  \citep{vecchia1988estimation,stein2004approximating,Guinness2018,Katzfuss_2020}, similar to the NNGP, where the ordering of the data is more complicated but results in a better approximation. These Vechias approximation techniques of ordering can be applied naturally in the proposed RNNC model, however, they are out of the scope of this paper and will be investigated in future work.  Next steps will include extending the proposed method in the  multivariate setting by using ideas from parallel partial autoregressive co-kriging \citep{ma2019multifidelity} and NNGP spatial factor models \citep{taylor2018spatial}.   Spherical covariance function can be used for global data set analysis \citep{GUINNESS_SphericalCov}, however their extension to anisotropic representation is not straightforward.  Still, work needs to be done in developing new strategies for tuning the hyperparameters in more complex covariance functions with multiple parameters within a fidelity level.

\section*{Acknowledgements}
The research of  Konomi and Kang  was supported in part by National Science Foundation grant NSF DMS-2053668 and the Taft Research Center at the University of Cincinnati. Kang was also  supported in part by  Simons Foundation's Collaboration Award (\#317298 and \#712755). 

\singlespacing \setlength{\bibsep}{5pt}

\bibliographystyle{jasa3}

\bibliography{reference1}

\begin{thebibliography}{45}
\newcommand{\enquote}[1]{``#1''}
\expandafter\ifx\csname natexlab\endcsname\relax\def\natexlab#1{#1}\fi
\expandafter\ifx\csname url\endcsname\relax
  \def\url#1{{\tt #1}}\fi
\expandafter\ifx\csname urlprefix\endcsname\relax\def\urlprefix{URL }\fi

\bibitem[\protect\citeauthoryear{Abdulah, Li, Cao, Ltaief, Keyes, Genton, and
  Sun}{Abdulah et~al.}{2023}]{Abdulah2023}
Abdulah, S., Li, Y., Cao, J., Ltaief, H., Keyes, D.~E., Genton, M.~G., and Sun,
  Y. (2023), \enquote{Large-scale environmental data science with ExaGeoStatR,}
  {\em Environmetrics\/}, 34, e2770,
  \urlprefix\url{https://onlinelibrary.wiley.com/doi/abs/10.1002/env.2770}.

\bibitem[\protect\citeauthoryear{Banerjee, Gelfand, Finley, and Sang}{Banerjee
  et~al.}{2008}]{banerjee2008gaussian}
Banerjee, S., Gelfand, A.~E., Finley, A.~O., and Sang, H. (2008),
  \enquote{Gaussian predictive process models for large spatial data sets,}
  {\em Journal of the Royal Statistical Society: Series B (Statistical
  Methodology)\/}, 70, 825--848.

\bibitem[\protect\citeauthoryear{Bayarri, Berger, and Liu}{Bayarri
  et~al.}{2009}]{Bayarri_module}
Bayarri, M.~J., Berger, J.~O., and Liu, F. (2009), \enquote{{Modularization in
  Bayesian analysis, with emphasis on analysis of computer models},} {\em
  Bayesian Analysis\/}, 4, 119 -- 150,
  \urlprefix\url{https://doi.org/10.1214/09-BA404}.

\bibitem[\protect\citeauthoryear{Chander, Hewison, Fox, Wu, Xiong, and
  Blackwell}{Chander et~al.}{2013}]{chander2013}
Chander, G., Hewison, T., Fox, N., Wu, X., Xiong, X., and Blackwell, W. (2013),
  \enquote{Overview of Intercalibration of Satellite Instruments,} {\em IEEE
  Transactions on Geoscience and Remote Sensing\/}, 51:3, 1056--1080.

\bibitem[\protect\citeauthoryear{Cheng, Konomi, Matthews, Karagiannis, and
  Kang}{Cheng et~al.}{2021}]{Si_etall2020_NNCGP}
Cheng, S., Konomi, B.~A., Matthews, J.~L., Karagiannis, G., and Kang, E.~L.
  (2021), \enquote{Hierarchical Bayesian nearest neighbor co-kriging Gaussian
  process models; an application to intersatellite calibration,} {\em Spatial
  Statistics\/}, 44, 100516,
  \urlprefix\url{https://www.sciencedirect.com/science/article/pii/S2211675321000269}.

\bibitem[\protect\citeauthoryear{Cressie and Johannesson}{Cressie and
  Johannesson}{2008}]{cressie2008fixed}
Cressie, N. and Johannesson, G. (2008), \enquote{Fixed rank kriging for very
  large spatial data sets,} {\em Journal of the Royal Statistical Society:
  Series B (Statistical Methodology)\/}, 70, 209--226.

\bibitem[\protect\citeauthoryear{Datta, Banerjee, Finley, and Gelfand}{Datta
  et~al.}{2016}]{datta2016hierarchical}
Datta, A., Banerjee, S., Finley, A.~O., and Gelfand, A.~E. (2016),
  \enquote{Hierarchical nearest-neighbor Gaussian process models for large
  geostatistical datasets,} {\em Journal of the American Statistical
  Association\/}, 111, 800--812.

\bibitem[\protect\citeauthoryear{Du, Zhang, Mandrekar et~al.}{Du
  et~al.}{2009}]{du2009fixed}
Du, J., Zhang, H., Mandrekar, V., et~al. (2009), \enquote{Fixed-domain
  asymptotic properties of tapered maximum likelihood estimators,} {\em the
  Annals of Statistics\/}, 37, 3330--3361.

\bibitem[\protect\citeauthoryear{Finley, Datta, Cook, Morton, Andersen, and
  Banerjee}{Finley et~al.}{2019}]{finley2019JCGS}
Finley, A.~O., Datta, A., Cook, B.~D., Morton, D.~C., Andersen, H.~E., and
  Banerjee, S. (2019), \enquote{Efficient Algorithms for Bayesian Nearest
  Neighbor Gaussian Processes,} {\em Journal of Computational and Graphical
  Statistics\/}, 28, 401--414,
  \urlprefix\url{https://doi.org/10.1080/10618600.2018.1537924}. PMID:
  31543693.

\bibitem[\protect\citeauthoryear{Furrer, Genton, and Nychka}{Furrer
  et~al.}{2006}]{furrer2006covariance}
Furrer, R., Genton, M.~G., and Nychka, D. (2006), \enquote{Covariance tapering
  for interpolation of large spatial datasets,} {\em Journal of Computational
  and Graphical Statistics\/}, 15, 502--523.

\bibitem[\protect\citeauthoryear{Gneiting and Raftery}{Gneiting and
  Raftery}{2007}]{GneitingRaftery2007}
Gneiting, T. and Raftery, A.~E. (2007), \enquote{Strictly Proper Scoring Rules,
  Prediction, and Estimation,} {\em Journal of the American Statistical
  Association\/}, 102, 359--378.

\bibitem[\protect\citeauthoryear{Goldberg}{Goldberg}{2011}]{goldberg2011}
Goldberg, M., e.~a. (2011), \enquote{The Global Space-Based Inter-Calibration
  Systems,} {\em Bull. Am. Meteorol. Soc.\/}, 92, 467--475.

\bibitem[\protect\citeauthoryear{Gramacy and Apley}{Gramacy and
  Apley}{2015}]{Gramacy2015}
Gramacy, R.~B. and Apley, D.~W. (2015), \enquote{Local {Gaussian} Process
  Approximation for Large Computer Experiments,} {\em Journal of Computational
  and Graphical Statistics\/}, 24, 561--578.

\bibitem[\protect\citeauthoryear{Guinness}{Guinness}{2018}]{Guinness2018}
Guinness, J. (2018), \enquote{Permutation and Grouping Methods for Sharpening
  Gaussian Process Approximations,} {\em Technometrics\/}, 60, 415--429,
  \urlprefix\url{https://doi.org/10.1080/00401706.2018.1437476}. PMID:
  31447491.

\bibitem[\protect\citeauthoryear{Guinness and Fuentes}{Guinness and
  Fuentes}{2016}]{GUINNESS_SphericalCov}
Guinness, J. and Fuentes, M. (2016), \enquote{Isotropic covariance functions on
  spheres: Some properties and modeling considerations,} {\em Journal of
  Multivariate Analysis\/}, 143, 143--152,
  \urlprefix\url{https://www.sciencedirect.com/science/article/pii/S0047259X15002109}.

\bibitem[\protect\citeauthoryear{Jackson, Wylie, and Bates}{Jackson
  et~al.}{2003}]{jackson2003}
Jackson, D., Wylie, D., and Bates, J. (2003), \enquote{The HIRS pathfinder
  radiance data set (1979–2001),} in {\em Proc.\ of the 12th Conference on
  Satellite Meteorology and Oceanography, Long Beach, CA, USA, 10-13 February
  2003\/}, volume 5805 of {\em LNCS\/}, Springer.

\bibitem[\protect\citeauthoryear{Katzfuss}{Katzfuss}{2017}]{Katzfuss2016}
Katzfuss, M. (2017), \enquote{A Multi-Resolution Approximation for Massive
  Spatial Datasets,} {\em Journal of the American Statistical Association\/},
  112, 201--214.

\bibitem[\protect\citeauthoryear{Katzfuss and Guinness}{Katzfuss and
  Guinness}{2021}]{katzfuss2017general}
Katzfuss, M. and Guinness, J. (2021), \enquote{{A General Framework for Vecchia
  Approximations of Gaussian Processes},} {\em Statistical Science\/}, 36, 124
  -- 141, \urlprefix\url{https://doi.org/10.1214/19-STS755}.

\bibitem[\protect\citeauthoryear{Katzfuss, Guinness, Gong, and Zilber}{Katzfuss
  et~al.}{2020}]{Katzfuss_2020}
Katzfuss, M., Guinness, J., Gong, W., and Zilber, D. (2020), \enquote{Vecchia
  Approximations of Gaussian-Process Predictions,} {\em Journal of
  Agricultural, Biological and Environmental Statistics\/}, 25, 383--414.

\bibitem[\protect\citeauthoryear{Kaufman, Schervish, and Nychka}{Kaufman
  et~al.}{2008}]{kaufman2008covariance}
Kaufman, C.~G., Schervish, M.~J., and Nychka, D.~W. (2008), \enquote{Covariance
  tapering for likelihood-based estimation in large spatial data sets,} {\em
  Journal of the American Statistical Association\/}, 103, 1545--1555.

\bibitem[\protect\citeauthoryear{Kennedy and O'Hagan}{Kennedy and
  O'Hagan}{2000}]{kennedy2000predicting}
Kennedy, M.~C. and O'Hagan, A. (2000), \enquote{Predicting the output from a
  complex computer code when fast approximations are available,} {\em
  Biometrika\/}, 87, 1--13.

\bibitem[\protect\citeauthoryear{Konomi, Hanandeh, Ma, and Kang}{Konomi
  et~al.}{2019}]{Konomi_env2019}
Konomi, B.~A., Hanandeh, A.~A., Ma, P., and Kang, E.~L. (2019),
  \enquote{Computationally efficient nonstationary nearest-neighbor Gaussian
  process models using data-driven techniques,} {\em Environmetrics\/}, 30,
  e2571,
  \urlprefix\url{https://onlinelibrary.wiley.com/doi/abs/10.1002/env.2571}.
  E2571 env.2571.

\bibitem[\protect\citeauthoryear{Konomi, Kang, Almomani, and Hobbs}{Konomi
  et~al.}{2023}]{Konomi2022_JABES}
Konomi, B.~A., Kang, E.~L., Almomani, A., and Hobbs, J. (2023),
  \enquote{Bayesian Latent Variable Co-kriging Model in Remote Sensing for
  Quality Flagged Observations,} {\em Journal of Agricultural, Biological and
  Environmental Statistics\/}, 4, 119 -- 150,
  \urlprefix\url{https://doi.org/10.1007/s13253-023-00530-9}.

\bibitem[\protect\citeauthoryear{Konomi and Karagiannis}{Konomi and
  Karagiannis}{2021}]{konomikaragiannisABTCK2019}
Konomi, B.~A. and Karagiannis, G. (2021), \enquote{Bayesian Analysis of
  Multifidelity Computer Models With Local Features and Nonnested Experimental
  Designs: Application to the WRF Model,} {\em Technometrics\/}, 63, 510--522,
  \urlprefix\url{https://doi.org/10.1080/00401706.2020.1855253}.

\bibitem[\protect\citeauthoryear{Le~Gratiet}{Le~Gratiet}{2013}]{le2013bayesian}
Le~Gratiet, L. (2013), \enquote{Bayesian analysis of hierarchical multifidelity
  codes,} {\em SIAM/ASA Journal on Uncertainty Quantification\/}, 1, 244--269.

\bibitem[\protect\citeauthoryear{Le~Gratiet and Garnier}{Le~Gratiet and
  Garnier}{2014}]{le2014recursive}
Le~Gratiet, L. and Garnier, J. (2014), \enquote{Recursive co-kriging model for
  design of computer experiments with multiple levels of fidelity,} {\em
  International Journal for Uncertainty Quantification\/}, 4.

\bibitem[\protect\citeauthoryear{Lindgren, Rue, and Lindstr{\"o}m}{Lindgren
  et~al.}{2011}]{lindgren2011explicit}
Lindgren, F., Rue, H., and Lindstr{\"o}m, J. (2011), \enquote{An explicit link
  between Gaussian fields and Gaussian Markov random fields: the stochastic
  partial differential equation approach,} {\em Journal of the Royal
  Statistical Society: Series B (Statistical Methodology)\/}, 73, 423--498.

\bibitem[\protect\citeauthoryear{Liu, Wong, and Kong}{Liu
  et~al.}{1994}]{LiuEtAl1994}
Liu, J.~S., Wong, W.~H., and Kong, A. (1994), \enquote{{Covariance structure of
  the Gibbs sampler with applications to the comparisons of estimators and
  augmentation schemes},} {\em Biometrika\/}, 81, 27--40,
  \urlprefix\url{https://doi.org/10.1093/biomet/81.1.27}.

\bibitem[\protect\citeauthoryear{Ma and Kang}{Ma and Kang}{2020}]{MaKang2020}
Ma, P. and Kang, E.~L. (2020), \enquote{A Fused Gaussian Process Model for Very
  Large Spatial Data,} {\em Journal of Computational and Graphical
  Statistics\/}, 29, 479--489.

\bibitem[\protect\citeauthoryear{Ma, Karagiannis, Konomi, Asher, Toro, and
  Cox}{Ma et~al.}{2022}]{ma2019multifidelity}
Ma, P., Karagiannis, G., Konomi, B.~A., Asher, T.~G., Toro, G.~R., and Cox,
  A.~T. (2022), \enquote{Multifidelity computer model emulation with
  high-dimensional output: An application to storm surge,} {\em Journal of the
  Royal Statistical Society: Series C (Applied Statistics)\/}, n/a,
  \urlprefix\url{https://rss.onlinelibrary.wiley.com/doi/abs/10.1111/rssc.12558}.

\bibitem[\protect\citeauthoryear{Matthew J.~Heaton and Terres}{Matthew
  J.~Heaton and Terres}{2017}]{Heaton_Tech2017}
Matthew J.~Heaton, W. F.~C. and Terres, M.~A. (2017), \enquote{Nonstationary
  Gaussian Process Models Using Spatial Hierarchical Clustering from Finite
  Differences,} {\em Technometrics\/}, 59, 93--101.

\bibitem[\protect\citeauthoryear{Michele~Peruzzi and Finley}{Michele~Peruzzi
  and Finley}{2022}]{MeshedGP2020}
Michele~Peruzzi, S.~B. and Finley, A.~O. (2022), \enquote{Highly Scalable
  Bayesian Geostatistical Modeling via Meshed Gaussian Processes on Partitioned
  Domains,} {\em Journal of the American Statistical Association\/}, 117,
  969--982.

\bibitem[\protect\citeauthoryear{{National Research Council}}{{National
  Research Council}}{2004}]{nrc2004}
{National Research Council} (2004), {\em Climate Data Records from
  Environmental Satellites: Interim Report\/}, Washington, DC: The National
  Academies Press,
  \urlprefix\url{https://www.nap.edu/catalog/10944/climate-data-records-from-environmental-satellites-interim-report}.

\bibitem[\protect\citeauthoryear{Nguyen, Cressie, and Braverman}{Nguyen
  et~al.}{2012}]{nguyen2012spatial}
Nguyen, H., Cressie, N., and Braverman, A. (2012), \enquote{Spatial statistical
  data fusion for remote sensing applications,} {\em Journal of the American
  Statistical Association\/}, 107, 1004--1018.

\bibitem[\protect\citeauthoryear{Nguyen, Cressie, and Braverman}{Nguyen
  et~al.}{2017}]{nguyen2017multivariate}
--- (2017), \enquote{Multivariate spatial data fusion for very large remote
  sensing datasets,} {\em Remote Sensing\/}, 9, 142.

\bibitem[\protect\citeauthoryear{Nychka, Bandyopadhyay, Hammerling, Lindgren,
  and Sain}{Nychka et~al.}{2015}]{nychka2015multiresolution}
Nychka, D., Bandyopadhyay, S., Hammerling, D., Lindgren, F., and Sain, S.
  (2015), \enquote{A multiresolution Gaussian process model for the analysis of
  large spatial datasets,} {\em Journal of Computational and Graphical
  Statistics\/}, 24, 579--599.

\bibitem[\protect\citeauthoryear{O’Hagan}{O’Hagan}{1998}]{o1998markov}
O’Hagan, A. (1998), \enquote{A Markov property for covariance structures,}
  {\em Statistics Research Report\/}, 98, 510.

\bibitem[\protect\citeauthoryear{Qian, Seepersad, Joseph, Allen, and
  Jeff~Wu}{Qian et~al.}{2005}]{ZhiguangConnerJanetWu2005}
Qian, Z., Seepersad, C.~C., Joseph, V.~R., Allen, J.~K., and Jeff~Wu, C.~F.
  (2005), \enquote{{Building Surrogate Models Based on Detailed and Approximate
  Simulations},} {\em Journal of Mechanical Design\/}, 128, 668--677,
  \urlprefix\url{https://doi.org/10.1115/1.2179459}.

\bibitem[\protect\citeauthoryear{Sang and Huang}{Sang and
  Huang}{2012}]{Sang2012}
Sang, H. and Huang, J.~Z. (2012), \enquote{{A full scale approximation of
  covariance functions for large spatial data sets},} {\em Journal of the Royal
  Statistical Society: Series B (Statistical Methodology)\/}, 74, 111--132.

\bibitem[\protect\citeauthoryear{Shirota, Finley, Cook, and Banerjee}{Shirota
  et~al.}{2023}]{Shirota_Finley2022}
Shirota, S., Finley, A.~O., Cook, B.~D., and Banerjee, S. (2023),
  \enquote{Conjugate sparse plus low rank models for efficient Bayesian
  interpolation of large spatial data,} {\em Environmetrics\/}, 34, e2748,
  \urlprefix\url{https://onlinelibrary.wiley.com/doi/abs/10.1002/env.2748}.

\bibitem[\protect\citeauthoryear{Stein}{Stein}{2014}]{stein2014limitations}
Stein, M.~L. (2014), \enquote{Limitations on low rank approximations for
  covariance matrices of spatial data,} {\em Spatial Statistics\/}, 8, 1--19.

\bibitem[\protect\citeauthoryear{Stein, Chi, and Welty}{Stein
  et~al.}{2004}]{stein2004approximating}
Stein, M.~L., Chi, Z., and Welty, L.~J. (2004), \enquote{Approximating
  likelihoods for large spatial data sets,} {\em Journal of the Royal
  Statistical Society: Series B (Statistical Methodology)\/}, 66, 275--296.

\bibitem[\protect\citeauthoryear{Taylor-Rodriguez, Finley, Datta, Babcock,
  Andersen, Cook, Morton, and Banerjee}{Taylor-Rodriguez
  et~al.}{2018}]{taylor2018spatial}
Taylor-Rodriguez, D., Finley, A.~O., Datta, A., Babcock, C., Andersen, H.-E.,
  Cook, B.~D., Morton, D.~C., and Banerjee, S. (2018), \enquote{Spatial Factor
  Models for High-Dimensional and Large Spatial Data: An Application in Forest
  Variable Mapping,} {\em arXiv preprint arXiv:1801.02078\/}.

\bibitem[\protect\citeauthoryear{Vecchia}{Vecchia}{1988}]{vecchia1988estimation}
Vecchia, A.~V. (1988), \enquote{Estimation and model identification for
  continuous spatial processes,} {\em Journal of the Royal Statistical Society:
  Series B (Methodological)\/}, 50, 297--312.

\bibitem[\protect\citeauthoryear{Xiong, Cao, and Chander}{Xiong
  et~al.}{2010}]{xiong2010}
Xiong, X., Cao, C., and Chander, G. (2010), \enquote{An overview of sensor
  calibration inter-comparison and applications,} {\em Frontiers of Earth
  Science in China\/}, 4, 237--252.

\end{thebibliography}

\appendix
\newpage
\section*{Appendix}

\section{NNGP specifications}\label{App1}

The posterior distribution of {\small{}{}{}
\begin{align}
\tilde{p}(\bfw_{t}|\cdot) & \propto\text{exp}\left[-\frac{1}{2}\sum_{i=1}^{n_{t}}\left\{ w_{t}(\bs_{t,i})-\mathbf{B}_{t,\bs_{t,i}}\bfw_{t,N_{t}(\bs_{t,i})}\right\} ^{T}F_{t,\bs_{t,i}}^{-1}\left\{ w_{t}(\bs_{t,i})-\mathbf{B}_{t,\bs_{t,i}}\bfw_{t,N_{t}(\bs_{t,i})}\right\} \right]\nonumber \\
 & =\text{exp}\left(-\frac{1}{2}\bfw_{t}^{T}\mathbf{B}_{t}^{T}\mathbf{F}_{t}^{-1}\mathbf{B}_{t}\bfw_{t}\right),
\end{align}
}{\small\par}
where $\mathbf{F}_{t}=\text{diag}(F_{t,\bs_{t,1}},F_{t,\bs_{t,2}},\ldots,F_{t,\bs_{t,n_{t}}})$,
$\mathbf{B}_{t}=\Big{(}\mathbf{B}_{t,1}^{T},\mathbf{B}_{t,2}^{T},\ldots,\mathbf{B}_{t,n_{t}}^{T}\Big{)}^{T}$,
and for each element in $\mathbf{B}_{t}$, we have $\mathbf{B}_{t,i}=\Big{(}\mathbf{B}_{t,s_{t,i},1}^{T},\mathbf{B}_{t,s_{t,i},2}^{T},\ldots,\mathbf{B}_{t,s_{t,i},n_{t}}^{T}\Big{)}^{T}$
and {\small{}{}{} 
\begin{align}
\mathbf{B}_{t,s_{t,i},j}=\begin{cases}
1,\ \text{if}\ i=j,\\
-\mathbf{B}_{t,s_{t,i}}[,k],\ \text{if}\ s_{t,j}\ \text{is the}\ k^{th}\ \text{element in}\ N_{t}(s_{t,i}),\\
0,\ \text{Others}.
\end{cases}
\end{align}
}

\section{Mean and Variance Specifications}

The mean vector $\boldsymbol{\mu} =  \left(\mu_1(\bs_{1,1}),\ldots,\mu_1(\bs_{1,n_1}),\ldots,\mu_T(\bs_{T,n_T})\right)$ is
\allowdisplaybreaks
\begin{align}
\mu_t(\bs_{t,k})=&\mathbf{1}_{\{t>1\}}(t)\sum_{i=1}^{t-1}\left\{\prod_{j=i}^{t-1}\zeta_{j}(\bs_{t,k})\right\} \left\{\mathbf{h}_{i}^{T}(\bs_{t,k})\boldsymbol{\beta}_{i} + \mathbf{1}_{\{\bs_{t,k}\in \bfS_i\}}(\bs_{t,k})w_i(\bs_{t,k}) \right\} \nonumber \\ &+\mathbf{h}_{t}^{T}(\bs_{t,k})\boldsymbol{\beta}_{t} + w_t(\bs_{t,k}),  \label{mean_function_original}
\end{align}
for $t=1,\ldots T$, $  i=1,\ldots, n_t$. $\mathbf{1}_{\{\cdot\}}(\cdot)$ is the indicator function,  and covariance matrix $\boldsymbol{\Lambda}$ is a block matrix with blocks $\Lambda^{(1,1)},\ldots,\Lambda^{(1,T)},\ldots,\Lambda^{(T,T)}$, and the size of $\boldsymbol{\Lambda}$ is $\sum_{t=1}^{T}n_t \times \sum_{t=1}^{T}n_t$. The $\Lambda^{(t,t)}$ 
components are calculated as:
\allowdisplaybreaks {
\begin{align}
 & \Lambda^{(t,t)}_{k,l} =  \text{cov}(z_{t}(\bs_{t,k}),z_{t}(\bs_{t,l})|\cdot)=\sum_{i=1}^{t-1} \mathbf{1}_{\{\bs_{t,k},\bs_{t,l}\notin \bfS_i\}}(\bs_{t,k},\bs_{t,l}) \left\{ \prod_{j=i}^{t-1}\zeta_{j}(\bs_{t,k})^{T}\zeta_{j}(\bs_{t,l})\right\} C_{i}(\bs_{t,k},\bs_{t,l}|\boldsymbol{\theta}_{i}) \nonumber \\
 & \qquad \qquad \qquad  +\mathbf{1}_{\bs_{t,k}=\bs_{t,l}}(\bs_{t,k},\bs_{t,l})\tau_{t}^{2},\nonumber 
 \end{align}
for $t \ \text{and}\ t' = 1,\ldots,T$; $k=1,\ldots,n_t$; $l = 1,\ldots,n_{t'}$,  and
 \begin{align}
 &\Lambda^{(t,t')}_{k,l} =  \text{cov}(z_{t}(s_{t,k}),z_{t'}(\bs_{t',l})|\cdot)=\sum_{i=1}^{\text{min}(t,t')-1} \mathbf{1}_{\{\bs_{t,k},\bs_{t',l}\notin \bfS_i\}}(\bs_{t,k},\bs_{t',l}) \left\{ \prod_{j=i}^{\text{min}(t,t')-1}\zeta_{j}(\bs_{t,k})^{T}\zeta_{j}(\bs_{t',l})\right\} \nonumber \\ 
 & \qquad \qquad \qquad  \times C_{i}(\bs_{t,k\bs_{t',l}}|\boldsymbol{\theta}_{i})  + \mathbf{1}_{\{\bs_{t,k},\bs_{t',l}\notin \bfS_{\text{min}(t,t')}\}}(\bs_{t,k},\bs_{t',l}) C_{\text{min}(t,t')}(\bs_{t,k},\bs_{t',l}|\boldsymbol{\theta}_{\text{min}(t,t')}),\label{covariance_function_original}
\end{align}
for $t\neq t'$, $\Lambda^{(t,t')}$. 
}
\section{Gibbs Sampler }
\begin{align}
\mathbf{V}_{\beta_t}^* &= (\mathbf{h}_t(\bfS_t)\tilde{\Lambda}_t(\bfS_t,\bftheta_{t})^{-1}\mathbf{h}^T_t(\bfS_t) + \mathbf{V}_{\beta_t}^{-1})^{-1},\nonumber \\
\bfmu_{\beta_t}^* & = \mathbf{V}_{\beta_t}^{-1}\bfmu_{\beta_t}+\mathbf{h}_t(\bfS_t)\tilde{\Lambda}_t(\bfS_t,\bftheta_{t})^{-1}(z_t(\bfS_t)-\zeta_{t-1}(\bfS_t)\hat{y}_{t-1}(\bfS_t)). \\
\mathbf{V}_{\gamma_{t}}^* &= \left[(\mathbf{g}_t^T(\bfS_{t+1})\hat{y}_{t}(\bfS_{t+1}))^T\tilde{\Lambda}_{t+1}(\bfS_{t+1},\bftheta_{t+1},\tau_{t+1})^{-1}(\mathbf{g}_t^T(\bfS_{t+1})\hat{y}_{t}(\bfS_{t+1})) + \mathbf{V}_{\gamma_{t}}^{-1}\right]^{-1},\nonumber \\
\bfmu_{\gamma_{t}}^* & = \mathbf{V}_{\gamma_{t}}^{-1}\bfmu_{\gamma_{t}}+(\mathbf{g}_t^T(\bfS_{t+1})\hat{y}_{t}(\bfS_{t+1}))^T\tilde{\Lambda}_{t+1}(\bfS_{t+1},\bftheta_{t+1},\tau_{t+1})^{-1}(\bfZ_{t+1}-\mathbf{h}_t^T(\bfS_{t+1})\bfbeta_{t+1}).\label{AppendixC11}
\end{align}

\section{Conjugate Conditional Probabilities}
we derive the posterior distribution as
\begin{align*}
     p(&\bfbeta_t,\bfgamma_{t-1},\sigma_t^2 |\bfZ_t,\hat{y}_{t-1}(\bfS_t))  \propto  IG(\sigma_t^2|a_t,b_t)N(\bfbeta_t|\bfmu_{\bfbeta_t},\sigma_t^2\mathbf{V}_{\bfbeta_t})N(\bfgamma_{t-1}|\bfmu_{\bfgamma_{t-1}},\sigma_t^2\mathbf{V}_{\bfgamma_{t-1}}) \nonumber \\
     & \quad \ \times N(\bfZ_t|\zeta_{t-1}(\bfS_t)\circ \hat{y}_{t-1}(\bfS_t)+\mathbf{h}^T_t\bfbeta_t,\sigma_t^2\tilde{\bfSigma}_t) \\
    & \propto p(\sigma_t^2|\bfZ_t,\hat{y}_{t-1}(\bfS_t))p(\bfbeta_t|\sigma_t^2,\bfZ_t,\hat{y}_{t-1}(\bfS_t))p(\bfgamma_{t-1}|\bfbeta_t,\sigma_t^2,\bfZ_t,\hat{y}_{t-1}(\bfS_t)) \nonumber \\
    & \propto (\sigma_t^2)^{a_t+0.5n_t}\text{exp}\left(-\frac{1}{2\sigma_t^2}(\bfbeta_t - \bfmu_{\beta_t})^T\mathbf{V}_{\beta_t}^{-1}(\bfbeta_t - \bfmu_{\beta_t})\right) \\
   & \quad \ \times \text{exp}\left(-\frac{1}{2\sigma_t^2}(\bfgamma_{t-1} - \bfmu_{\gamma_{t-1}})^T\mathbf{V}_{\gamma_{t-1}}^{-1}(\bfgamma_{t-1} - \bfmu_{\gamma_{t-1}})\right) \\
   & \quad \ \times \text{exp}\left(-\frac{1}{2\sigma_t^2}(\bfZ_t - \mathbf{g}^T(\bfS_t)\bfgamma_{t-1} \hat{y}_{t-1}(\bfS_t)-\mathbf{h}^T_t(\bfS_t)\bfbeta_t)^T \tilde{\bfSigma}_t^{-1}(\bfZ_t - \mathbf{g}^T(\bfS_t)\bfgamma_{t-1} \hat{y}_{t-1}(\bfS_t)-\mathbf{h}^T_t(\bfS_t)\bfbeta_t) \right).
 \end{align*}
 
 The full conditional density function of $\bfgamma_{t-1}$ is 
\begin{align}
p(\bfgamma_{t-1}&|\bfbeta_t,\sigma_t^2,\bfZ_t,\hat{y}_{t-1}(\bfS_t)) \propto \text{exp}\Bigg(-\frac{1}{2\sigma_t^2}[\mathbf{g}(\bfS_t)\bfgamma_{t-1}^T\hat{y}_{t-1}(\bfS_t)^T\tilde{\bfSigma}_t^{-1}\mathbf{g}^T(\bfS_t)\bfgamma_{t-1} \hat{y}_{t-1}(\bfS_t) \nonumber \\
& - 2(\bfZ_t-\mathbf{h}^T_t(\bfS_t)\bfbeta_t)^T\tilde{\Sigma}_t^{-1}\mathbf{g}^T(\bfS_t)\bfgamma_{t-1} \hat{y}_{t-1}(\bfS_t) ]\Bigg) \nonumber  \\
&\propto N(\bfgamma_{t-1}|\tilde{\mathbf{V}}_{\bfgamma_{t-1}}\tilde{\bfmu}_{\bfgamma_{t-1}},\sigma^2\tilde{\mathbf{V}}_{\bfgamma_{t-1}}), \nonumber  \\
\tilde{\bfmu}_{\bfgamma_{t-1}}& =  \mathbf{V}_{\bfgamma_{t-1}}^{-1}\bfmu_{\bfgamma_{t-1}} +\mathbf{g}(\bfS_t) \hat{y}_{t-1}(\bfS_t)^T\tilde{\Sigma}_t^{-1}(\bfZ_t-\mathbf{h}^T_t(\bfS_t)\bfbeta_t), \nonumber \\
\tilde{\mathbf{V}}_{\bfgamma_{t-1}} & = \left( \mathbf{V}_{\bfgamma_{t-1}}^{-1} + \mathbf{g}(\bfS_t)\hat{y}_{t-1}(\bfS_t)^T\tilde{\bfSigma}_t^{-1}\hat{y}_{t-1}(\bfS_t)\mathbf{g}^T(\bfS_t) \right)^{-1}. \label{conjugate_gamma_b}
\end{align}

After integrate $\bfgamma_{t-1}$ out, the conditional posterior density function of $\bfbeta_t$ is
\begin{align}
    p(\bfbeta_{t}&|\sigma_t^2,\bfZ_t,\hat{y}_{t-1}(\bfS_t))\propto \text{exp}\Bigg(-\frac{1}{2\sigma_t^2}[(\mathbf{h}^T_t(\bfS_t)\bfbeta_t)^T\tilde{\bfSigma}_t^{-1}(\mathbf{h}^T_t(\bfS_t)\bfbeta_t)-2\bfZ_t^T\tilde{\bfSigma}_t^{-1}\mathbf{h}^T_t(\bfS_t)\bfbeta_t]\Bigg) \nonumber  \\
   & \quad \  \times \text{exp}\Bigg( -\frac{1}{2\sigma_t^2}(\bfbeta_t-\bfmu_{\beta_t})^T\mathbf{V}_{\beta_t}^{-1}(\bfbeta_t-\bfmu_{\beta_t}) \Bigg) \text{exp}\Bigg(\frac{1}{2\sigma_t^2}\tilde{\bfmu}_{\bfgamma_{t-1}}^T\tilde{\mathbf{V}}_{\bfgamma_{t-1}}\tilde{\bfmu}_{\bfgamma_{t-1}} \Bigg), \nonumber  \\
   & \propto N(\bfbeta_t|\tilde{\mathbf{V}}_{\beta_t}\tilde{\bfmu}_{\beta_t}, \sigma_t^2\tilde{\mathbf{V}}_{\beta_t}), \nonumber \\
   \tilde{\bfmu}_{\beta_t} & = \mathbf{V}_{\beta_t}^{-1}\bfmu_{\beta_t} +\mathbf{h}_t(\bfS_t)\tilde{\bfSigma}_t^{-1}\bfZ_t - (\mathbf{g}(\bfS_t)y_{t-1}(\bfS_t)^T\tilde{\bfSigma}_t^{-1}\mathbf{h}^T_t(\bfS_t))^T \tilde{\mathbf{V}}_{\gamma_{t-1}}(\mathbf{V}_{\gamma_{t-1}}^{-1}\bfmu_{\gamma_{t-1}}\nonumber \\
   & \quad \  +\mathbf{g}(\bfS_t)\hat{y}_{t-1}(\bfS_t)^T\tilde{\bfSigma}_t^{-1}\bfZ_t), \nonumber  \\
   \tilde{\mathbf{V}}_{\beta_t} & = \Bigg( \mathbf{V}_{\beta_t}^{-1} + \mathbf{h}(\bfS_t)\tilde{\bfSigma}_t^{-1}\mathbf{h}(\bfS_t)^T - (\mathbf{g}(\bfS_t)\hat{y}_{t-1}(\bfS_t)^T\tilde{\bfSigma}_t^{-1}\mathbf{h}^T_t(\bfS_t))^T \tilde{\mathbf{V}}_{\bfgamma_{t-1}} (\mathbf{g}(\bfS_t)y_{t-1}(\bfS_t)^T\tilde{\bfSigma}_t^{-1}\mathbf{h}^T_t(\bfS_t)) \Bigg)^{-1}. \label{conjugate_beta5} 
\end{align}

The marginalized posterior density function of $\sigma_t$ is
\begin{align}
    &p(\sigma_t^2|\bfZ_t,\hat{y}_{t-1}(\bfS_t)) \propto \sigma_t^{-a_t-0.5n_t}\text{exp}\Bigg( -\frac{1}{2\sigma_t^2}\Bigg[2b_t + \bfZ_t^T\tilde{\bfSigma}_t^{-1}\bfZ_t+ \bfmu_{\beta_t}^T\mathbf{V}_{\beta_t}^{-1}\bfmu_{\beta_t} +\bfmu_{\gamma_{t-1}}^T\mathbf{V}_{\gamma_{t-1}}^{-1}\bfmu_{\gamma_{t-1}} - \tilde{\bfmu}_{\beta_t}^T\tilde{\mathbf{V}}_{\beta_t}\tilde{\bfmu}_{\beta_t}  \nonumber \\
    & \qquad  - (\mathbf{V}_{\bfgamma_{t-1}}^{-1}\bfmu_{\bfgamma_{t-1}} +\mathbf{g}(\bfS_t) \hat{y}_{t-1}(\bfS_t)^T\tilde{\bfSigma}_t^{-1}\bfZ_t)^T \tilde{\mathbf{V}}_{\bfgamma_{t-1}} (\mathbf{V}_{\bfgamma_{t-1}}^{-1}\bfmu_{\bfgamma_{t-1}} +\mathbf{g}(\bfS_t) \hat{y}_{t-1}(\bfS_t)^T\tilde{\bfSigma}_t^{-1}\bfZ_t)\Bigg] \Bigg),  \nonumber \\
    & \sigma_t^2|\bfZ_t,\hat{y}_{t-1}(\bfS_t) \sim IG(\sigma_t^2|a_t^*, b_t^*), \nonumber  \\
    & a_t^* = a_t + n_t/2, \nonumber  \\
    & b_t^* = b_t + 0.5\Bigg(\bfZ_t^T\tilde{\bfSigma}_t^{-1}\bfZ_t+\bfmu_{\beta_t}^T\mathbf{V}_{\beta_t}^{-1}\bfmu_{\beta_t} + \bfmu_{\gamma_{t-1}}^T\mathbf{V}_{\gamma_{t-1}}^{-1}\bfmu_{\gamma_{t-1}} - \tilde{\bfmu}_{\beta_t}^T\tilde{\mathbf{V}}_{\beta_t}\tilde{\bfmu}_{\beta_t} \nonumber \\
    & \qquad  - (\mathbf{V}_{\bfgamma_{t-1}}^{-1}\bfmu_{\bfgamma_{t-1}} +\mathbf{g}(\bfS_t) \hat{y}_{t-1}(\bfS_t)^T\tilde{\bfSigma}_t^{-1}\bfZ_t)^T \tilde{\mathbf{V}}_{\bfgamma_{t-1}} (\mathbf{V}_{\bfgamma_{t-1}}^{-1}\bfmu_{\bfgamma_{t-1}} +\mathbf{g}(\bfS_t) \hat{y}_{t-1}(\bfS_t)^T\tilde{\bfSigma}_t^{-1}\bfZ_t)\Bigg). \label{conjugate_sigma_3}
\end{align}

the conditional posterior density function of $\bfbeta_1$ is
\begin{align}
        \bfbeta_{1}|\sigma_1^2,\bfZ_1 & \sim N(\bfbeta_1|\tilde{\mathbf{V}}_{\beta_1}\tilde{\bfmu}_{\beta_1}, \sigma_1^2\tilde{\mathbf{V}}_{\beta_1}), \nonumber \\
   \tilde{\bfmu}_{\beta_1} & = \mathbf{V}_{\beta_1}^{-1}\bfmu_{\beta_1} +\mathbf{h}_1(\bfS_1)\tilde{\bfSigma}_1^{-1}\bfZ_1, \nonumber  \\
   \tilde{\mathbf{V}}_{\beta_1} & = \Bigg( \mathbf{V}_{\beta_1}^{-1} + \mathbf{h}(\bfS_1)\tilde{\bfSigma}_1^{-1}\mathbf{h}(\bfS_1)^T \Bigg)^{-1}, \label{conjugate_beta1}
\end{align}
and the marginal posterior density function of $\sigma_1^2$ is 
\begin{align}
     \sigma_1^2|\bfZ_1 & \sim IG(\sigma_1^2|a_1^*, b_1^*), \nonumber  \\
     a_1^* & = a_1 + n_1/2, \nonumber  \\
    b_1^*  &= b_1 + 0.5\left(\bfZ_1^T\tilde{\bfSigma}_t^{-1}\bfZ_1 +\bfmu_{\beta_1}^T\mathbf{V}_{\beta_1}^{-1}\bfmu_{\beta_1} - \tilde{\bfmu}_{\beta_1}^T\tilde{\mathbf{V}}_{\beta_1}\tilde{\bfmu}_{\beta_1}\right). \label{conjugate_sigma_4}
\end{align}

\section{Performance Metrics}\label{Metrics}

In the empirical comparisons, we used the following performance metrics: 
\begin{enumerate}
\item Root mean square prediction error (RMSPE) is defined as 
\[
\text{RMSPE}=\sqrt{\frac{1}{n}\sum_{i=1}^{n}(y_{i}^{\text{pred}}-y_{i}^{\text{obs}})^{2}}
\]
 where $y^{\text{obs}}$ is the observed value in test data-set and
$y_{i}^{\text{pred}}$ is the predicted value from the model. It measures
the accuracy of the prediction from model. Smaller values of RMSPE
indicate more a accurate model.
\item Nash-Sutcliffe model efficiency coefficient (NSME) is defined
as: 
\begin{gather*}
\text{NSME}=1-\frac{\sum_{i=1}^{n}(y_{i}^{\text{pred}}-y_{i}^{\text{obs}})^{2}}{\sum_{i=1}^{n}(y_{i}^{\text{obs}}-\Bar{y}^{\text{obs}})^{2}}
\end{gather*}
where $y^{\text{obs}}$ is the observed value in test data-set and
$y_{i}^{\text{pred}}$ is the predicted value from the model. NSME
gives the relative magnitude of the residual variance from data and
the model variance. NSME values closer to $1$ indicate that the model has
a better predictive performance. 
\item 95\% CVG is the coverage probability of 95\% equal tail prediction
interval. 95\% CVG values closer to $0.95$ indicate better prediction
performance for the model. 
\item 95\% ALCI is average length of 95\% equal tail prediction interval.
Smaller 95\% ALCI values indicate better prediction performance for
the model. 
\item Continuous Ranked Probability Score (CRPS), which is defined as
\begin{align*}
\text{CRPS}(G,y) = \int (G-H_y)^2,    
\end{align*}
where $y$ is a scalar quantity that needs to forecast, and y admits an underlying distribution that is described by CDF $F$; $G$ is a CDF that is chosen by the forecaster in order to predict $F$. $H$ is a unit step function and $H_y(x)$ indicates a centered Heaviside function $H(x-y)$. CRPS is negatively oriented and the lower scores imply better performance.
\end{enumerate}

\newpage
\doublespacing

\section{Simulation Study with Four Fidelity Levels}
We consider a system with four levels of fidelity represented by the hierarchical
statistical model  (3.1) defined  on a two dimensional unit
square domain with univariate observation data sets  $\bfZ=(\bfZ_{1},\bfZ_{2},\bfZ_{3},\bfZ_{4})$ with corresponding spatial locations $\bfS=(\bfS_{1},\bfS_{2},\bfS_{3},\bfS_{4})$. We assume a constant mean component for each level with design matrix $\mathbf{h}(\bfS_{t})=\mathbbm{1}$ and autoregressive coefficient function to be constant as $\gamma_1=1.1$, $\gamma_2=0.9$, $\gamma_3=1$. We chose isotropic  exponential covariance function for the Gaussian process latent variables with parameters $(\sigma_1^2=2,\phi_1=12)$, $(\sigma_2^2=1,\phi_2=6)$, $(\sigma_3^2=0.8,\phi_3=8)$, and $(\sigma_4^2=0.5, \phi_4=3)$. Each of the levels are contaminated with white noise with variance $\tau_1=0.1$, $\tau_2=0.05$, $\tau_1=0.05$, and $\tau_4=0.01$. To make possible simulation of a Gaussian process with the specifications given in model (3.1), we constrain our overall sample size to $n=12,000$ and each level of fidelity $n_1=n_2=n_3=n_4=3,000$.

\begin{figure}
\centering
\subfloat[First level fidelity observations\label{fig:full1}]{\centering \includegraphics[width=0.49\linewidth]{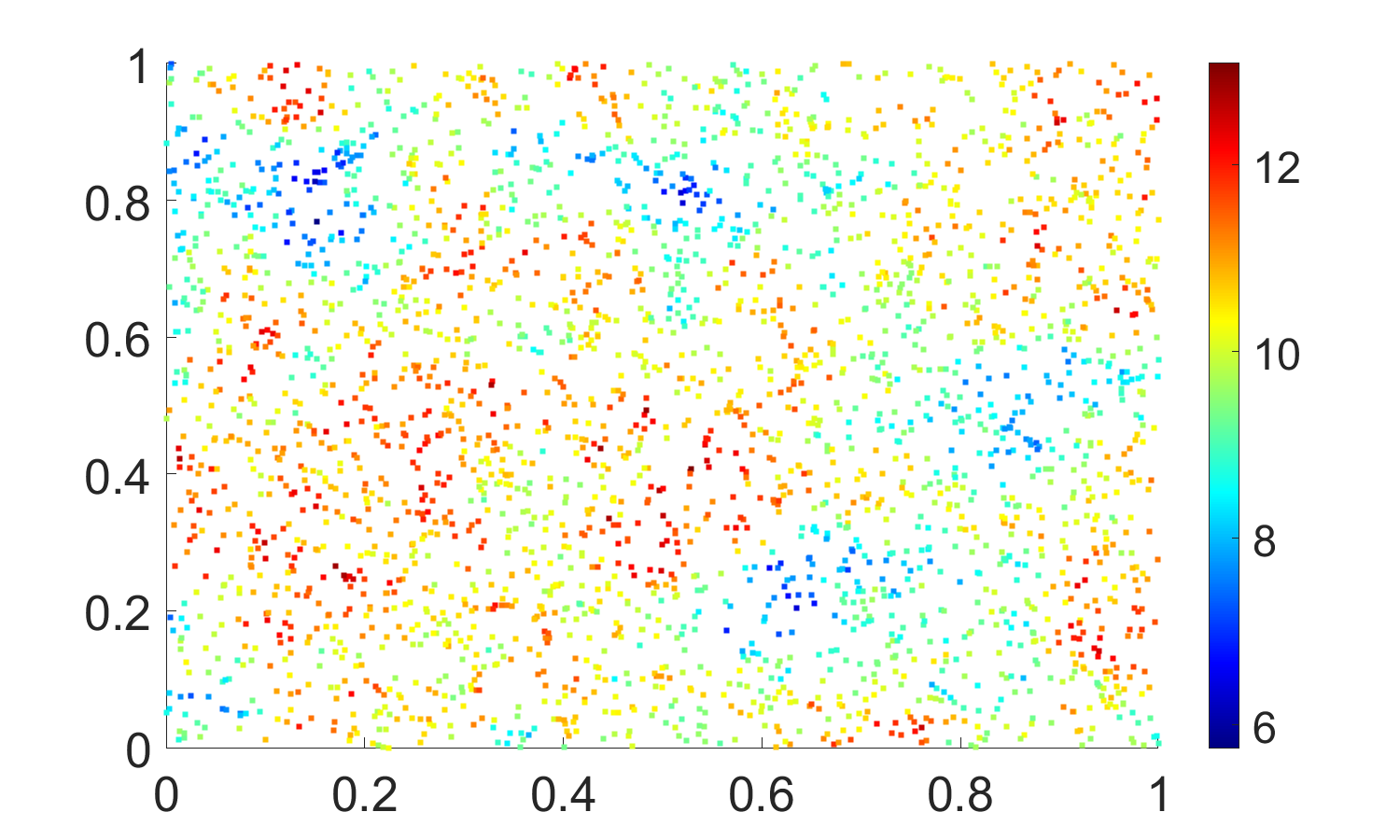}
}
\subfloat[Second level fidelity observations\label{fig:full2}]{\centering \includegraphics[width=0.49\linewidth]{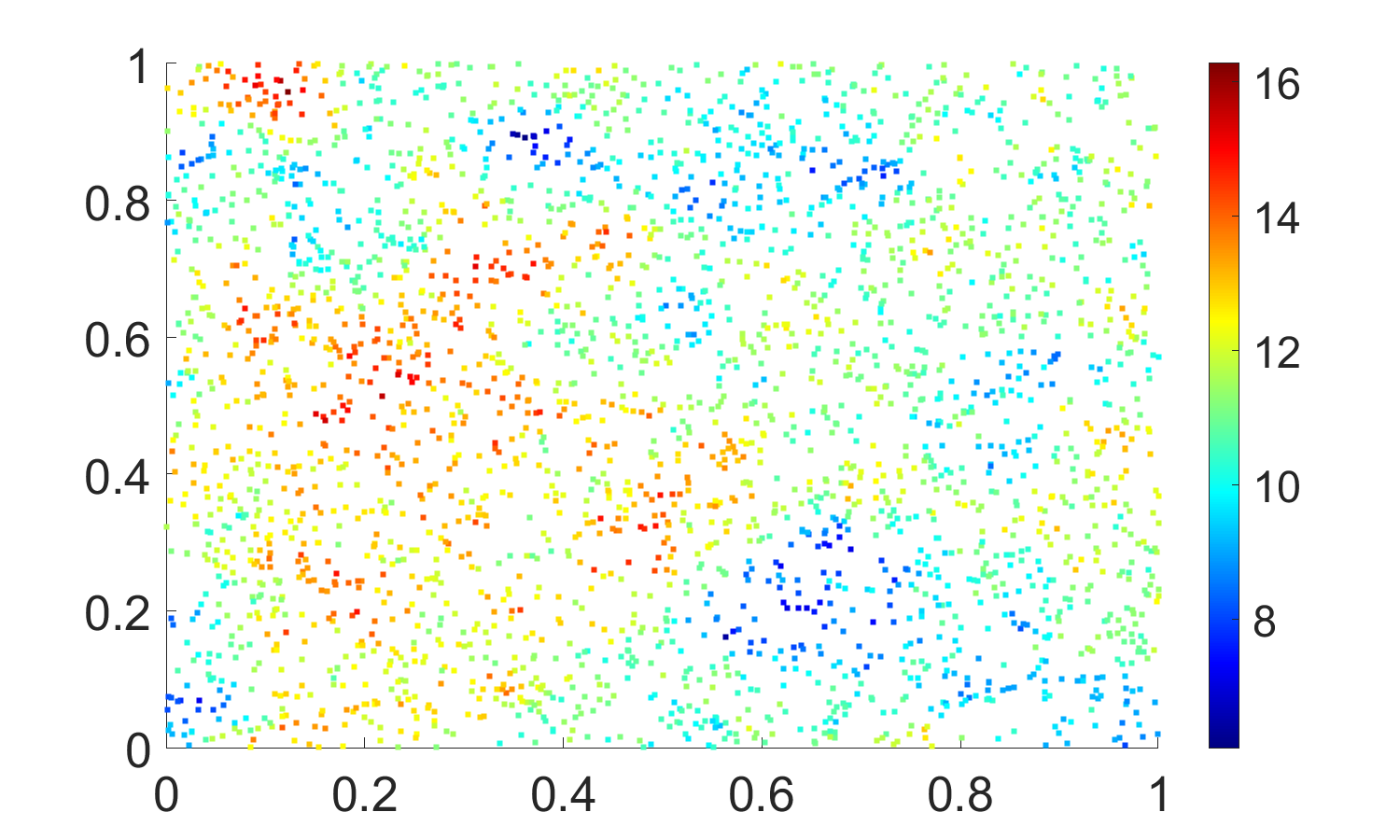}
}
\\
\subfloat[Third level fidelity observations\label{fig:full3}]{\centering \includegraphics[width=0.49\linewidth]{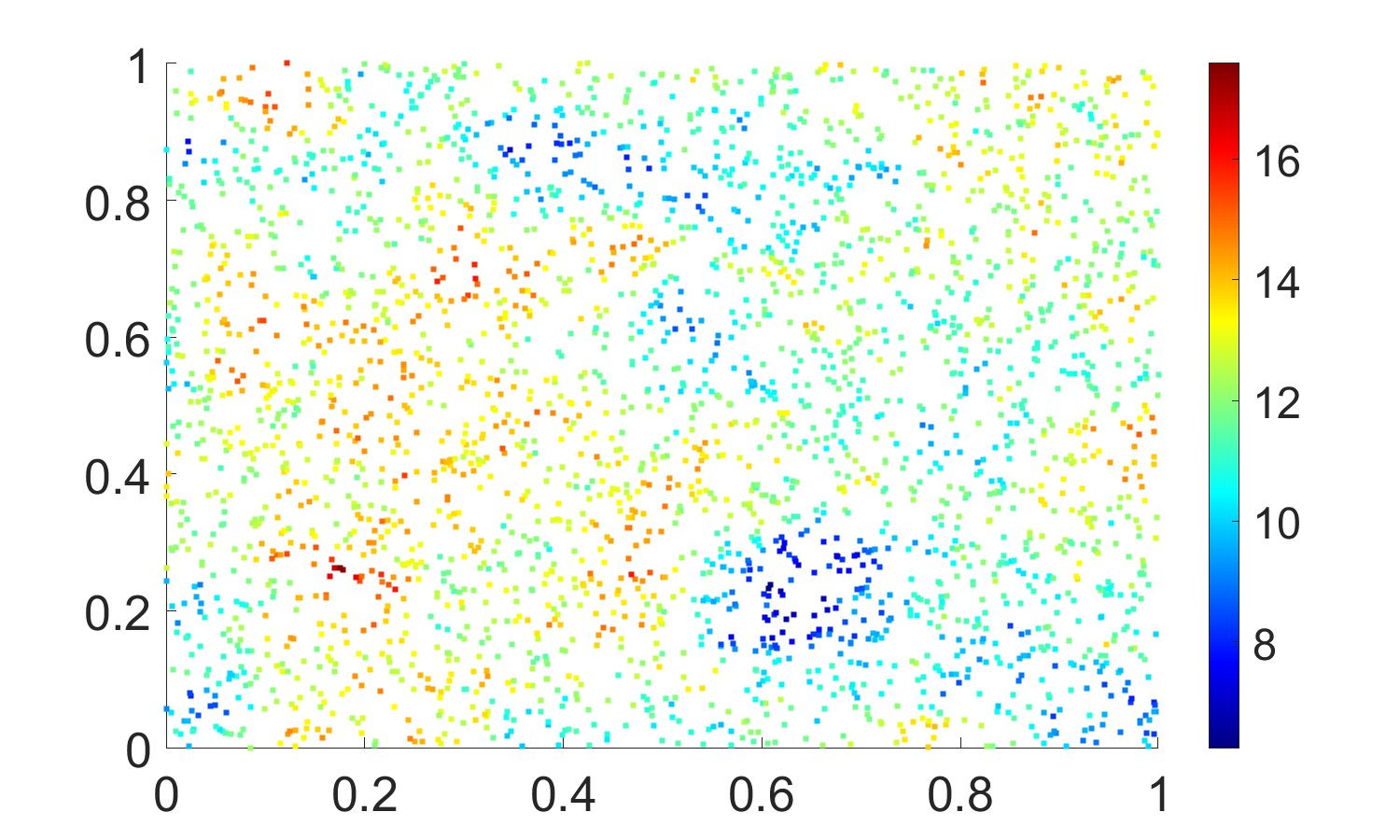}
}
\subfloat[Fourth level fidelity observations\label{fig:full4}]{\centering \includegraphics[width=0.49\linewidth]{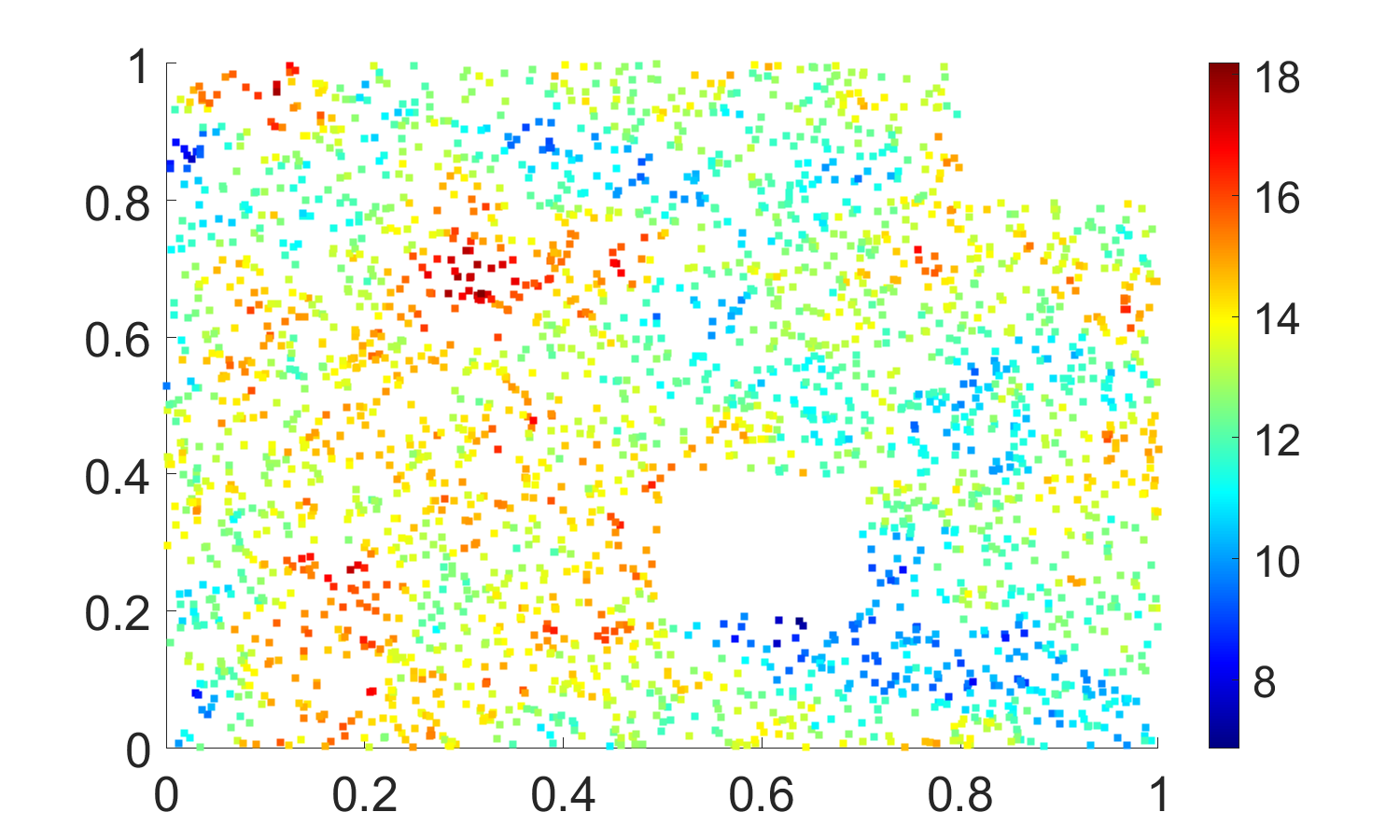}
}
\\
\subfloat[Fourth level fidelity testing data\label{fig:testing4}]{\centering \includegraphics[width=0.49\linewidth]{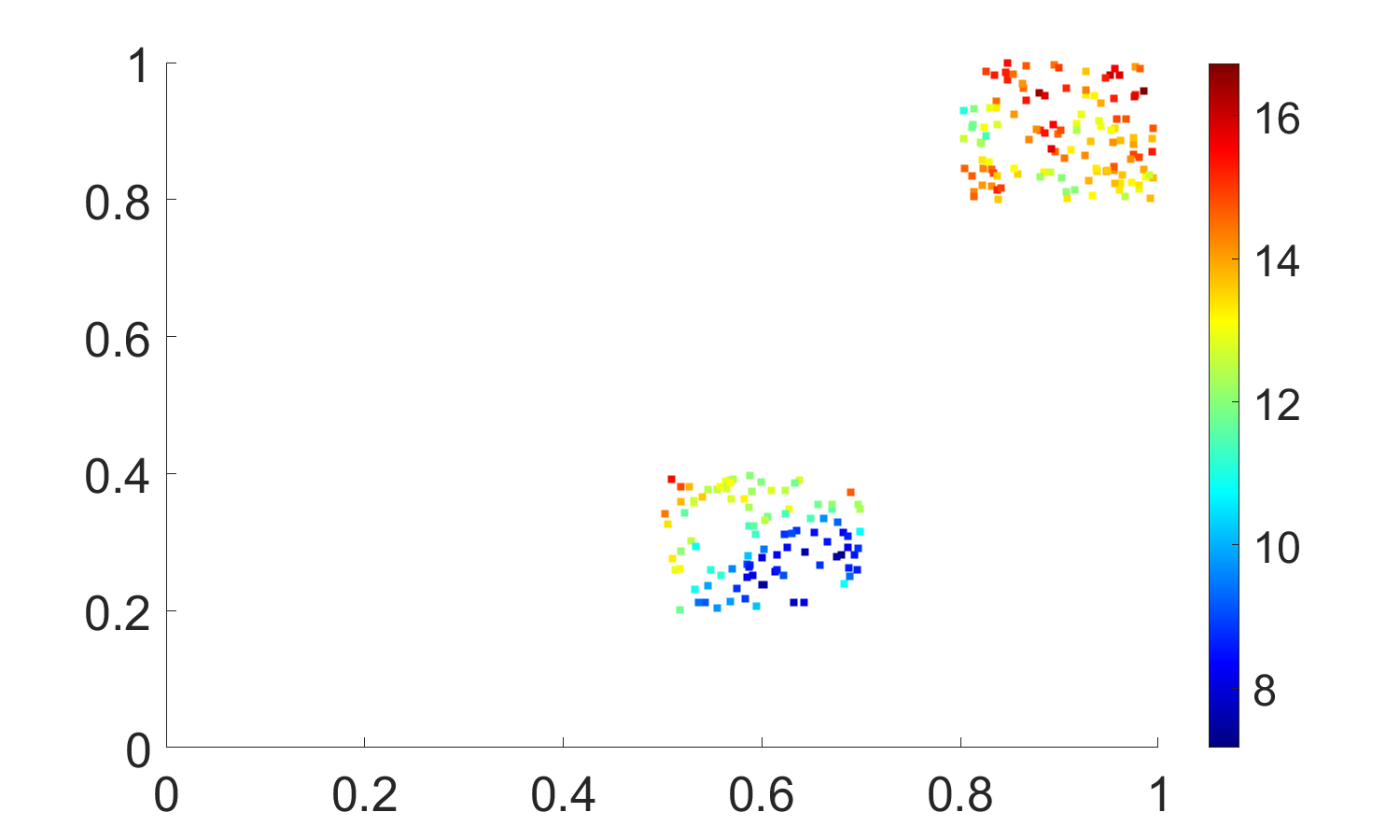}
}
\caption{Observations for four fidelity
level structure: a) first fidelity level training data, b) second fidelity level training data, c) third fidelity level training data, d)   forth  fidelity level training data (white boxes indicate the testing regions), e)  fourth fidelity level testing data. {\large{}{}\label{fig2Sup}}}
\end{figure}

We generate a synthetic data set for the above statistical model.  The data sets, shown in Figures \ref{fig:full1}, \ref{fig:full2}, \ref{fig:full3}, \ref{fig:full4} show the training data set for each of the four fidelity levels  and \ref{fig:testing4} show the testing data from the fourth fidelity level. We apply the conjugate RNNC model to the generated data sets.  We assign independent conjugate prior on parameters $\beta_{t} \sim N(0,1000)$ for $t=1,\dots, 4$, and scale parameter $\gamma_j\sim N(0,1000)$ for $j=1,2,3$. We assign independent inverse gamma prior on spatial variance parameters $\sigma_t^2\sim IG(2,1)$, and on the noise parameters  $\tau_t^2\sim IG(2,1)$ for $t=1, \dots, 4$. We also assign uniform prior on the range parameters $\phi_{1,t}\equiv\phi_{2,t} \sim U(0,30)$ for $t=1,\dots, 4$. The RMSPE with  a 5-fold cross-validation was used for the conjugate RNNC model.  We select $(\phi_t,\tilde{\tau}_t^2)$ on a grid such that $\phi_t$  ranges at   $[0.1 , 25]$ and $\tilde{\tau}_t^2$ ranges at $[0.0005 , 0.4]$.  As in the case of two fidelity levels, no significant differences were observed when we used 3-fold cross-validation and 7-fold cross-validation approach. 

\begin{figure}
\centering
\subfloat[Prediction Mean conjugate NNGP\label{fig:predS}]{\centering \includegraphics[width=0.49\linewidth]{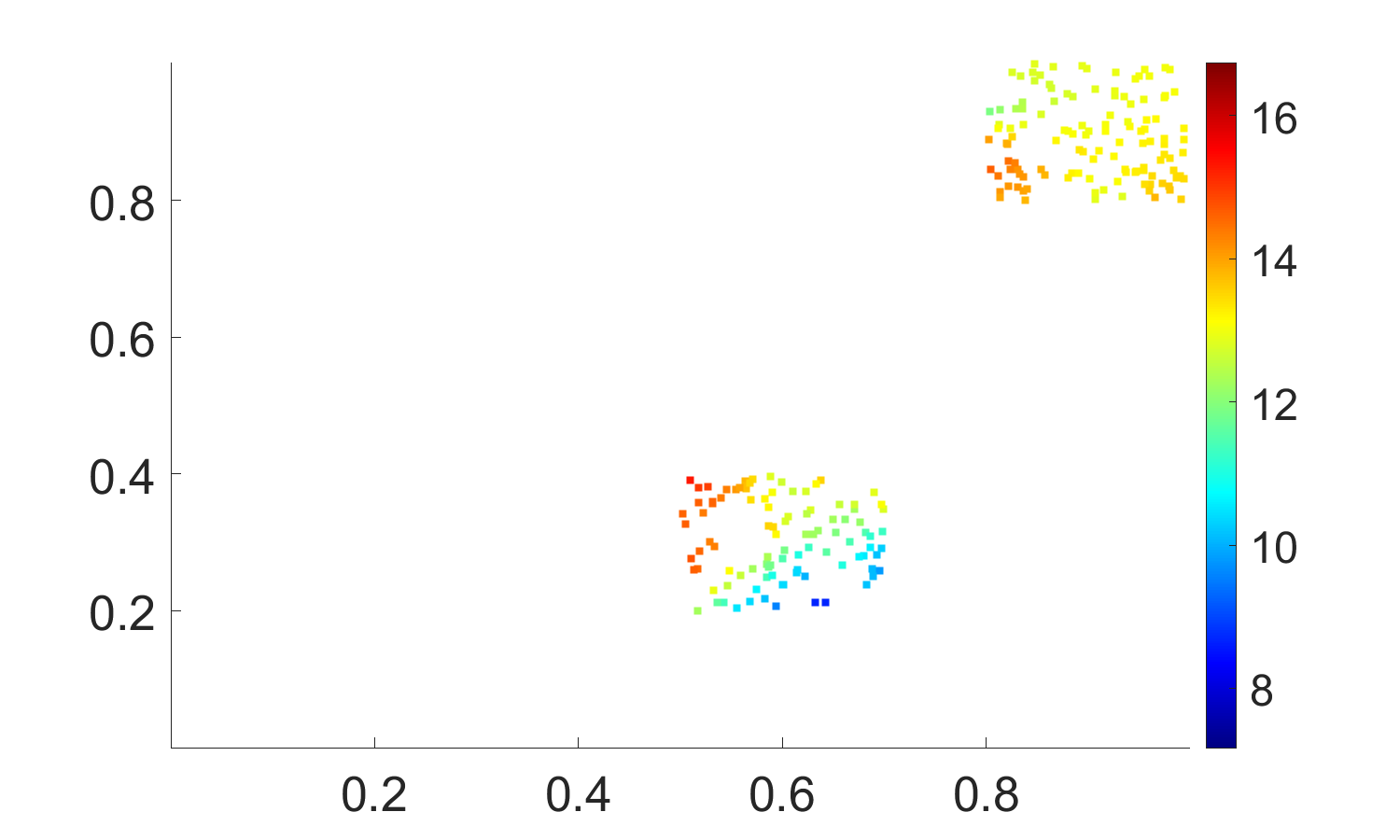}}
\subfloat[Prediction sd conjugate NNGP\label{fig:predicVarS}]{\centering \includegraphics[width=0.49\linewidth]{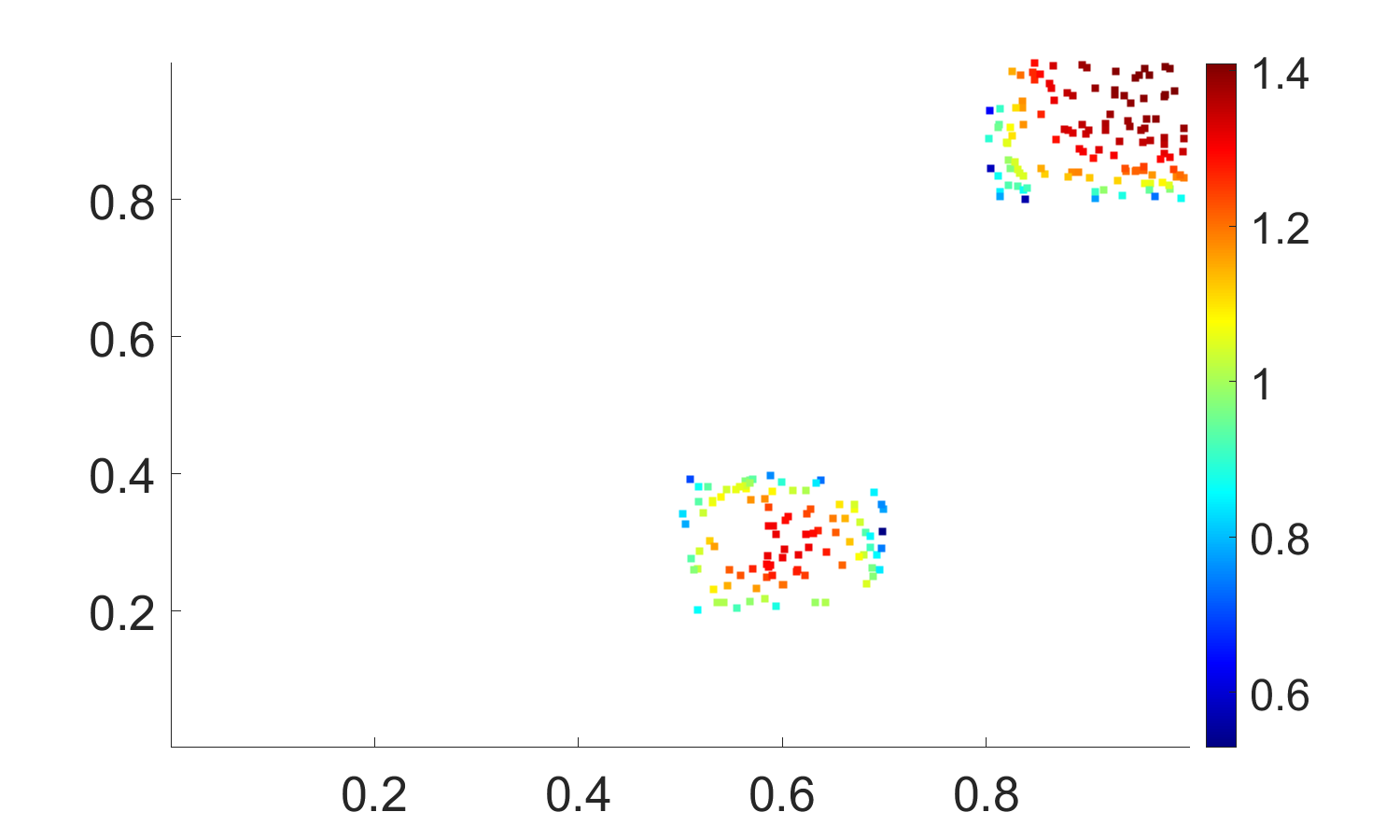}}
\\
\subfloat[Prediction Mean conjugate RNNC\label{fig:pred4}]{\centering \includegraphics[width=0.49\linewidth]{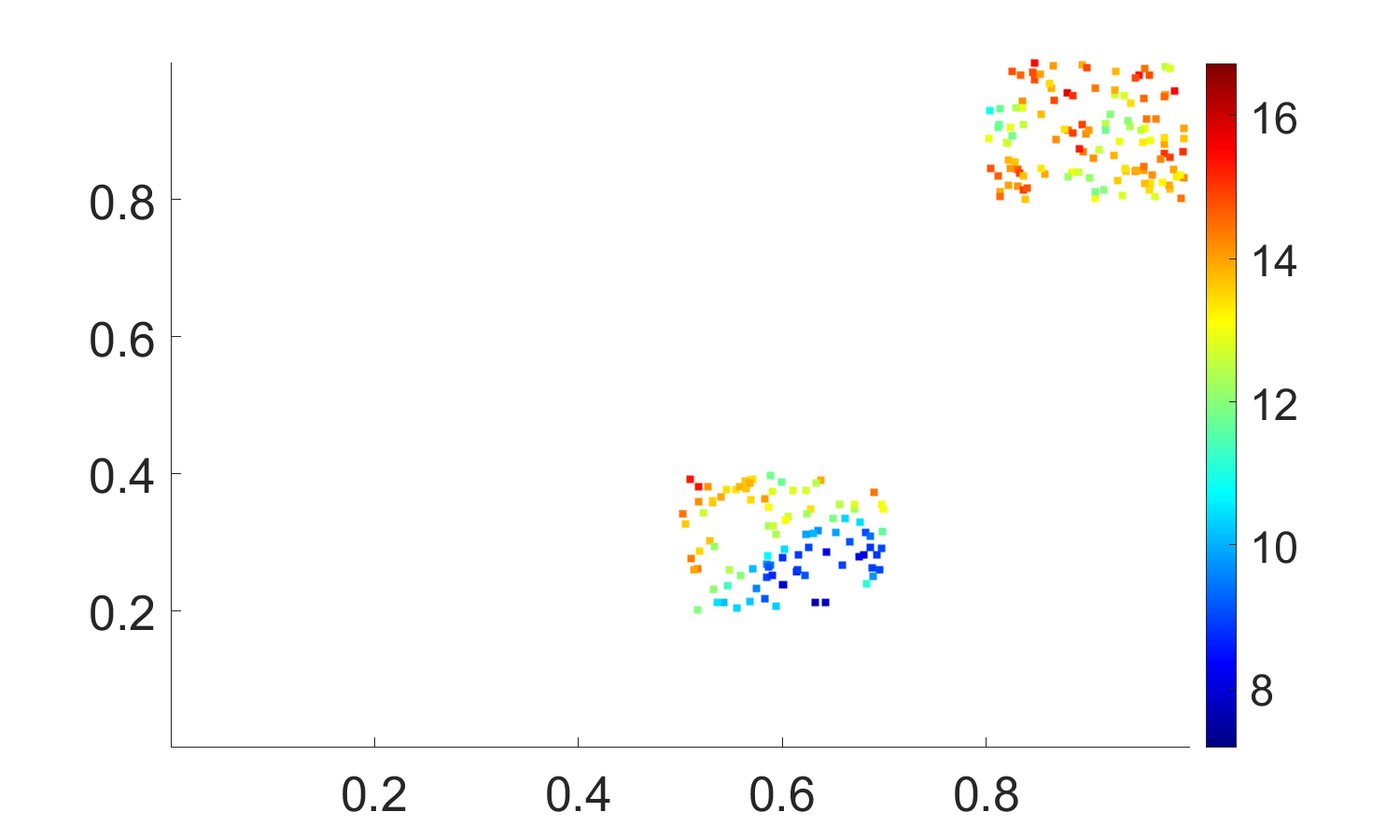}
}
\subfloat[Prediction sd conjugate RNNC\label{fig:predicVar}]{\centering \includegraphics[width=0.49\linewidth]{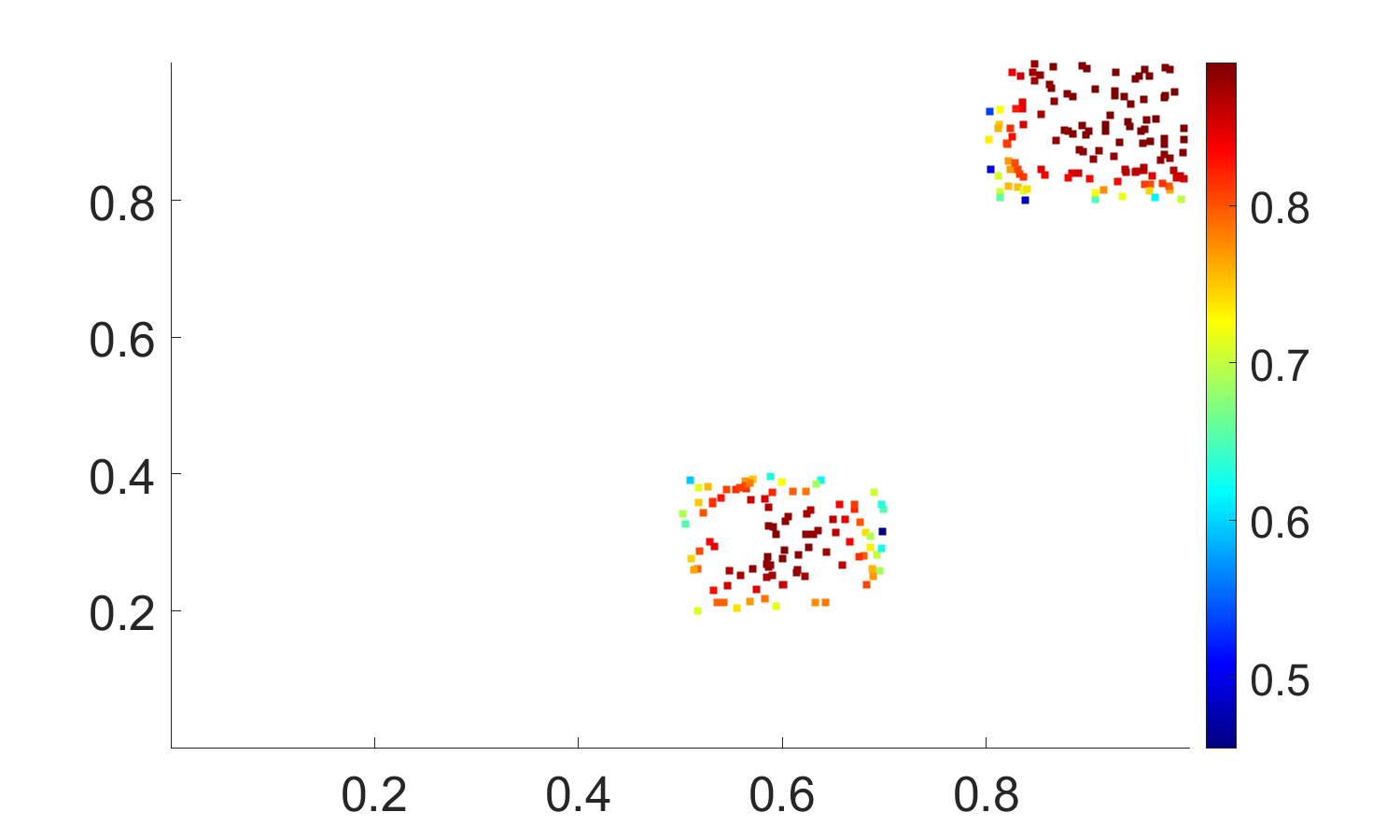}}
\caption{Prediction mean and the prediction standard deviations using both methods: a) prediction mean using conjugate NNGP, b) Prediction standard deviation (sd) using conjugate NNGP, c)  prediction mean using conjugate RNNC, and d)  Prediction standard deviation (sd) using conjugate RNNC. {\large{}{}\label{fig:pred}}}
\end{figure}

We compare our conjugate RNNC with the conjugate NNGP proposed in  \citep{finley2019JCGS} using only the fourth single level training data. The prior specification for the conjugate NNGP are the same as above. Figure~\ref{fig:pred} gives the prediction mean and the prediction standard deviations using both methods. Figure~\ref{fig:predS} gives the single layer conjugate NNGP and compared to our proposed method Figure~\ref{fig:pred} is less accurate. Figure~\ref{fig:predicVarS} shows the single layer conjugate NNGP  prediction standard deviations which is considerably larger than the prediction standard deviations using conjugate RNNC. Moreover, the computed conjugate RNNC RMSPE is $0.64$ with 95\% CVG $0.94$ and ALCI  $2.37$. Instead, the computed conjugate NNGP RMSPE is $1.68$ with 95\% CVG $0.86$ and ALCI  $3.98$. These comparisons shows that accounting for the multi-fidelity dependencies can  improve our prediction accuracy and its variation. The computational time for our proposed method is around four times slower than the conjugate NNGP in a single level. More precisely our method take around $117$ sec and the single layer conjugate NNGP $27$ seconds). This is very normal to expect since the computational complexity is linear to the observed data. 

\end{document}